\documentclass[prl,reprint,superscriptaddress,amsmath,amssymb,aps,longbibliography,floatfix]{revtex4-2}

\usepackage{bm}
\usepackage{mathtools}
\usepackage{microtype}
\usepackage{xcolor}
\usepackage{graphicx}
\graphicspath{{figures/}}
\usepackage{braket}
\usepackage{booktabs}
\usepackage{tikz}
\usetikzlibrary{arrows.meta,positioning,calc}
\usepackage{hyperref}
\hypersetup{colorlinks=true,citecolor=blue,linkcolor=blue,urlcolor=blue}
\usepackage{soul}

\newcommand{\ii}{\mathrm{i}}
\newcommand{\ee}{\mathrm{e}}
\newcommand{\cN}{\mathcal{N}}
\newcommand{\cO}{\mathcal{O}}

\definecolor{vacuumblue}{RGB}{0,114,178}
\definecolor{domainred}{RGB}{213,94,0}
\definecolor{kinkpurple}{RGB}{112,67,137}
\definecolor{bondgreen}{RGB}{0,128,104}

\begin{document}

\title{Discrete Scale Invariance of Ising Mesons}

\author{Denis V. Vasilyev}
\email{denis.vasilyev@imfp.org.cn}
\affiliation{Institute for Mathematics and Fundamental Physics, Hefei 230088, China}
\affiliation{Hefei National Laboratory, Hefei 230088, China}
\author{Abbas Ali Saberi}
\email{asaberi@constructor.university}
\affiliation{School of Science, Constructor University, Campus Ring 1, 28759 Bremen, Germany}

\date{\today}

\begin{abstract}
Discrete scale invariance often arises from resonant few-body or singular
scale-invariant dynamics. We show that it can instead emerge in the
internal motion of a composite excitation in an ordered quantum magnet.
In the marginal $1/r^4$ transverse-field Ising chain, long-range bonds
generate an asymptotic inverse-square attraction between dressed domain
walls, with strength fixed by the spontaneous magnetization. The dressed
two-kink threshold curvature sets the relative kinetic scale. Together,
these independently accessible quantities determine the scale-anomaly
exponent and hence the meson hierarchy. In the supercritical regime, the
same exponent governs geometric binding-energy and size ratios,
logarithmic level accumulation, and log-periodic threshold scattering.
Full-spin exact diagonalization and a fourth-order weak-field Hamiltonian
support the coupled energy--size scaling in the finite window between
core and ring effects. This provides a microscopic many-body realization
of quantum limit-cycle physics, with its universal scaling fixed
by magnetic order and kink mobility.
\end{abstract}

\maketitle

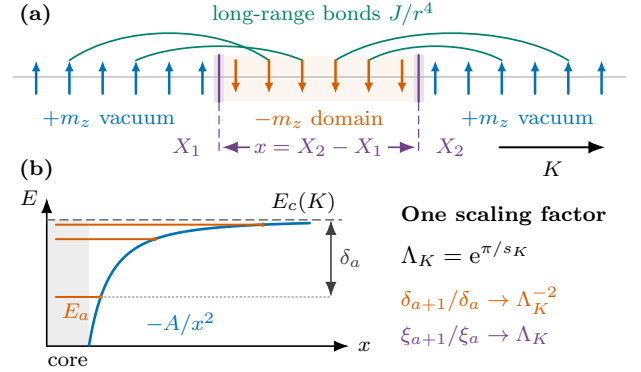
\begin{figure}[t]
\centering
\begin{tikzpicture}[
  x=1cm,y=1cm,
  >=Latex,
  line cap=round,line join=round,
  font=\fontsize{8}{9}\selectfont,
  spinup/.style={
    vacuumblue,line width=0.95pt,
    -{Latex[length=1.45mm,width=1.2mm]}
  },
  spindown/.style={
    domainred,line width=0.95pt,
    -{Latex[length=1.45mm,width=1.2mm]}
  },
  bond/.style={bondgreen,line width=0.65pt},
  guide/.style={
    kinkpurple,densely dashed,line width=0.45pt
  },
  level/.style={domainred,line width=0.75pt},
  panel/.style={
    anchor=west,
    font=\bfseries\fontsize{8}{9}\selectfont
  }
]
\path[use as bounding box]
  (-0.02,-0.12) rectangle (8.48,4.78);

\node[panel] at (0.08,4.56) {(a)};
\node[text=bondgreen] at (4.22,4.56)
  {long-range bonds $J/r^4$};

\fill[domainred!6]
  (2.84,3.42) rectangle (5.48,4.01);
\foreach \xwall in {2.84,5.48}{
  \fill[kinkpurple!16]
    (\xwall-0.065,3.40) rectangle (\xwall+0.065,4.02);
}

\draw[bond]
  (0.86,3.95)
  .. controls (1.55,4.43) and (2.81,4.43) ..
  (3.50,3.95);
\draw[bond]
  (1.74,3.95)
  .. controls (2.30,4.20) and (3.38,4.20) ..
  (3.94,3.95);
\draw[bond]
  (4.38,3.95)
  .. controls (5.15,4.43) and (6.69,4.43) ..
  (7.46,3.95);
\draw[bond]
  (4.82,3.95)
  .. controls (5.27,4.20) and (6.13,4.20) ..
  (6.58,3.95);

\draw[black!25,line width=0.55pt]
  (0.12,3.73)--(8.22,3.73);

\foreach \i in {0,1,2,3,4,5,12,13,14,15,16,17}{
  \pgfmathsetmacro{\xp}{0.42+0.44*\i}
  \draw[spinup] (\xp,3.51)--(\xp,3.95);
}
\foreach \i in {6,...,11}{
  \pgfmathsetmacro{\xp}{0.42+0.44*\i}
  \draw[spindown] (\xp,3.95)--(\xp,3.51);
}

\foreach \xwall in {2.84,5.48}{
  \draw[kinkpurple,line width=0.85pt]
    (\xwall,3.40)--(\xwall,4.02);
}

\node[text=vacuumblue] at (1.38,3.17)
  {$+m_z$ vacuum};
\node[text=domainred] at (4.16,3.17)
  {$-m_z$ domain};
\node[text=vacuumblue] at (6.92,3.17)
  {$+m_z$ vacuum};

\foreach \xwall in {2.84,5.48}{
  \draw[guide] (\xwall,3.39)--(\xwall,2.70);
}
\draw[<->,kinkpurple,line width=0.65pt]
  (2.88,2.80)--(5.44,2.80)
  node[midway,fill=white,inner sep=1.5pt]
  {$x=X_2-X_1$};

\node[text=kinkpurple,anchor=east] at (2.73,2.80)
  {$X_1$};
\node[text=kinkpurple,anchor=west] at (5.59,2.80)
  {$X_2$};

\draw[->,line width=0.75pt]
  (6.55,2.80)--(7.95,2.80)
  node[midway,below=2pt] {$K$};

\node[panel] at (0.08,2.51) {(b)};

\begin{scope}[shift={(0.56,0.17)}]
  \def\Ec{1.67}
  \def\Atail{0.52}
  \def\xcore{0.56}
  \def\dFirst{1.02}

  \fill[black!7] (0,0) rectangle (\xcore,\Ec);

  \draw[->,line width=0.6pt]
    (0,0)--(4.00,0) node[right] {$x$};
  \draw[->,line width=0.6pt]
    (0,0)--(0,1.96) node[left] {$E$};

  \draw[densely dashed,black!60,line width=0.55pt]
    (0,\Ec)--(3.90,\Ec);
  \node[anchor=south east,inner sep=1pt]
    at (3.87,1.72) {$E_c(K)$};

  \draw[
    vacuumblue,line width=1.0pt,
    domain=0.56:3.48,samples=100
  ]
    plot (\x,{\Ec-\Atail/(\x*\x)});

  \foreach \n in {0,1,2}{
    \pgfmathsetmacro{\binding}{\dFirst/pow(4,\n)}
    \pgfmathsetmacro{\ylevel}{\Ec-\binding}
    \pgfmathsetmacro{\xturn}{sqrt(\Atail/\binding)}

    \draw[level] (0.12,\ylevel)--(\xturn,\ylevel);
    \fill[domainred] (\xturn,\ylevel) circle[radius=0.8pt];
  }

  \pgfmathsetmacro{\yFirst}{\Ec-\dFirst}
  \pgfmathsetmacro{\xFirst}{sqrt(\Atail/\dFirst)}

  \node[
    text=domainred,anchor=north west,
    inner sep=1pt
  ] at (0.16,\yFirst-0.04) {$E_a$};

  \draw[densely dotted,black!40,line width=0.5pt]
    (\xFirst,\yFirst)--(3.76,\yFirst);
  \draw[<->,black!75,line width=0.6pt]
    (3.76,\yFirst)--(3.76,\Ec)
    node[midway,right=2pt,inner sep=1pt] {$\delta_a$};

  \node[anchor=north,inner sep=1pt]
    at (0.28,-0.10) {core};
  \node[text=vacuumblue,inner sep=1pt]
    at (1.77,0.29) {$-A/x^2$};
\end{scope}

\node[
  anchor=west,
  font=\bfseries\fontsize{8}{9}\selectfont
] at (5.13,1.9) {One scaling factor};

\node[
  anchor=west,
  font=\fontsize{9}{10}\selectfont
] at (5.13,1.4)
  {$\Lambda_K=\mathrm{e}^{\pi/s_K}$};
%

\node[anchor=west,text=domainred] at (5.13,0.78)
  {$\delta_{a+1}/\delta_a\to\Lambda_K^{-2}$};

\node[anchor=west,text=kinkpurple] at (5.13,0.31)
  {$\xi_{a+1}/\xi_a\to\Lambda_K$};

\end{tikzpicture}

\caption{From long-range bonds to geometric mesons.
(a)~A reversed domain is bounded by dressed walls at $X_1$ and $X_2$.
Their separation $x$ is the internal coordinate, while $K$ is the
total momentum. Arcs illustrate long-range bonds crossing either
interface; arrows indicate the local order.
(b)~The attractive tail $-A/x^2$ supports a geometric sequence of
bound states accumulating below the two-kink threshold $E_c(K)$
when $g_K>1/4$, Eq.~\eqref{eq:central}. Level segments end at their outer classical turning
points, illustrating the growing spatial scale of shallower states.
The same factor $\Lambda_K$ governs the binding energies
$\delta_a=E_c(K)-E_a(K)$ and meson sizes $\xi_a$.
The core and levels are schematic.}
\label{fig:model_schematic}
\end{figure}

\emph{Introduction.---}
Geometric hierarchies of bound states reveal a distinctive form of
universality: short-distance physics fixes an overall scale, while the
ratios of successive binding energies and sizes are set by long-distance
dynamics. Each successive state is larger by a fixed factor and less
tightly bound by its square. This discrete scale invariance underlies the
Efimov effect~\cite{Efimov1970,BraatenHammer2006}. For emergent particles
in quantum matter, linking the scaling factor to independently measurable
properties of their host would connect collective order to universal laws
for internal excitation spectra.

An attractive inverse-square potential provides the basic mechanism:
the potential and quadratic kinetic energy scale identically under
dilation~\cite{Case1950,Frank1971,EssinGriffiths2006}. Their balance
thus sets no intrinsic length, leaving no analogue of the Bohr radius. Above a critical dimensionless attraction, short-distance matching introduces an overall scale and reduces the continuous scaling symmetry of the asymptotic equation to a discrete one. This yields a geometric
bound-state spectrum and log-periodic scattering~\cite{Camblong2000,Beane2001,BawinCoon2003,MuellerHo2004,HammerSwingle2006,MorozSchmidt2010}.
Related scaling has been proposed for three-magnon states in quantum
magnets~\cite{Nishida2013} and observed in charged-vacancy resonances in
graphene~\cite{Ovdat2017}. Recent long-range spin-chain constructions use
a two-magnon resonance leading to three-magnon Efimov states or an
impurity-mediated few-body mechanism~\cite{SunFengZhang2026,SunFengZhangImpurity2026}.

Ising chains provide a distinct route to discrete scale invariance through the relative motion of a bound pair of domain walls. A longitudinal field confines domain walls
into mesons in the short-range Ising model~\cite{McCoyWu1978,FonsecaZamolodchikov2006,Rutkevich2005,Rutkevich2008},
with the exceptional $E_8$ spectrum emerging near criticality~\cite{Zamolodchikov1989,Coldea2010}. Confinement also shapes
relaxation~\cite{Kormos2017,James2019,Robinson2019,Lerose2020},
collisions~\cite{SuraceLerose2021,Karpov2022,Bennewitz2025},
and string breaking~\cite{Verdel2020,Surace2026}.
String-breaking dynamics have been observed in trapped-ion and
Rydberg-array simulators~\cite{De2024String,GonzalezCuadra2025String}.
Long-range Ising bonds, Fig.~\ref{fig:model_schematic}(a), bind the walls of a reversed interval while preserving the degeneracy of
the ordered vacua~\cite{Liu2019,Defenu2019,Vovrosh2022}. For interactions
$J/r^\alpha$ with $\alpha>2$, the pair approaches a finite dissociation
threshold without a bulk string tension. Summing the bonds gives the
known bare wall--wall attraction proportional to $-x^{2-\alpha}$~\cite{Liu2019}; at $\alpha=4$, it has precisely the inverse-square form.

The challenge is to establish this hierarchy for dressed quantum walls:
fluctuations renormalize both their interaction and mobility, while core
and finite-size effects restrict the observable scaling window. Here we
develop a microscopic theory in which the spontaneous magnetization $m_z$
fixes the asymptotic attraction amplitude, $A=2Jm_z^2/3$, and the dressed two-kink
threshold curvature $D_K$ fixes the relative kinetic energy at total
momentum $K$. For an isolated quadratic threshold with $D_K>0$, these
independently accessible quantities determine
\begin{equation}
g_K=\frac{2Jm_z^2}{3D_K},\qquad
s_K=\sqrt{g_K-\frac14},
\label{eq:central}
\end{equation}
with a geometric hierarchy for $g_K>1/4$, Fig.~\ref{fig:model_schematic}(b). The lattice fixes the
short-distance matching phase, while the ratio of magnetic order to kink
mobility fixes the scaling factor.

{We test the predicted energy and size ratios using full-spin exact
diagonalization and a fourth-order weak-field Hamiltonian.
Suppressing the quartic threshold dispersion enlarges the accessible
scaling window.} Logarithmic
level counting and log-periodic threshold scattering provide further
predictions of the same $s_K$ exponent. The resulting scale anomaly acts in the
internal coordinate of a two-kink composite while the bulk remains gapped
and ordered, establishing a quantitative connection between magnetic order,
emergent-particle motion, and universal excitation structure.

\emph{Model and marginal meson interaction.---}
We consider the ferromagnetic long-range transverse-field Ising chain
\cite{Vanderstraeten2018,Liu2019,DefenuRMP2023},
\begin{equation}
H=-J\sum_{i<j}\frac{\sigma_i^z\sigma_j^z}{r_{ij}^{\alpha}}
-h\sum_i\sigma_i^x,\qquad J,h>0,
\label{eq:H}
\end{equation}
where $\sigma_i^\mu$ are Pauli matrices and the lattice spacing is unity.
We work in the gapped ordered phase, whose two vacua have
$\langle\sigma_i^z\rangle_\pm=\pm m_z$. {A reversed domain therefore incurs no bulk energy cost:
its energy is set by the two interfaces and their interaction.}
The finite-ring couplings and numerical methods are specified in
Appendix~\ref{app:ED} of the Supplemental Material \cite{SupplementalMaterial}. For broader context on bulk long-range quantum criticality, see Ref.~\cite{Li2026QuantumCriticality}; here we focus on two-kink threshold physics within the gapped ordered phase.

\begin{figure}[!tbp]
\centering
\includegraphics[width=\columnwidth]{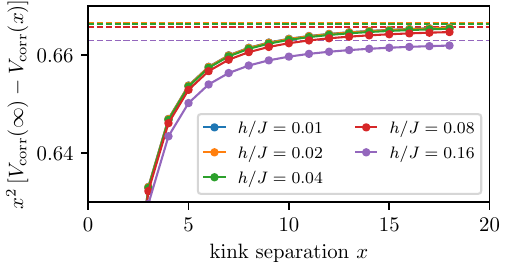}
\caption{Correlation reconstruction at $L=36$ for the indicated fields.
In units of $J$, symbols show
$x^2[V_{\rm corr}(\infty)-V_{\rm corr}(x)]$ and dashed lines show
$2Jm_z^2/3$. Correlations beyond $L/2$ are continued by their measured
plateau value, which fixes the asymptote. The nontrivial finite-$x$
approach from below follows the leading lattice correction derived in
Appendix~\ref{app:tail}.}
\label{fig:microscopic}
\end{figure}

A reversed domain of length $x$ cuts $2\min(r,x)$ bonds of range $r$
[Fig.~\ref{fig:model_schematic}(a)]. Far from the wall cores, cluster
decomposition fixes the spin product on each crossing bond to $-m_z^2$
rather than $+m_z^2$ in a uniform vacuum. Their difference fixes the leading
interaction by the same geometry as classical bond counting
(Appendix~\ref{app:tail}). The attraction approaches a finite limit as
$-x^{2-\alpha}$ for $\alpha>2$. At $\alpha=4$, assuming
power-decaying subleading corrections,
\begin{equation}
V_{\rm rel}(x)-V_{\rm rel}(\infty)
=-\frac{A}{x^2}+\cO(x^{-2-\omega}),\quad
A=\frac{2Jm_z^2}{3},
\label{eq:tail}
\end{equation}
where $\omega>0$ and $m_z$ is the fully dressed order parameter. Localized
interface structure changes the core and subleading terms but not the
coefficient fixed by long bonds joining the two ordered bulk regions.
{Under these spatial assumptions,} the argument is nonperturbative in~$h$.
Figure~\ref{fig:microscopic} provides a supporting
correlation reconstruction $V_{\rm corr}$ of the crossing-bond energy. The measured correlation plateau supplies
the asymptotic coefficient, while the finite-$x$ approach from below
{includes the plateau contribution's leading lattice correction (Appendix~\ref{app:tail}).}

Let $E_2(K,q)$ be the energy of two distant dressed kinks with total
momentum $K$ and relative momentum $q$.
For an isolated, nondegenerate minimum at $q=0$,
$E_2(K,q)=E_c(K)+D_Kq^2+C_Kq^4+\cO(q^6)$, with $D_K>0$.
The curvature $D_K$ measures the kinetic cost of relative motion.
It is obtained from a dressed one-kink band in a twisted ring, separately
from the meson spectrum; $m_z^2$ follows from the long-distance periodic
ground-state correlations (Appendix~\ref{app:ED}).
Eliminating virtual states outside this threshold channel
\cite{Feshbach1958,Feshbach1962} gives
\begin{equation}
H_{\rm rel}^{(K)}-E_c(K)
=-D_K\partial_x^2-\frac{A}{x^2}+W_K.
\label{eq:relative}
\end{equation}
This reduction requires an isolated two-kink threshold in the
chosen symmetry sector. At weak field, the finite gap to additional
kink pairs supports the single-channel description away from
channel crossings; an open eliminated continuum requires a
multichannel or resonance treatment.
Here $W_K$ collects core terms, subleading interactions, and higher
gradients. For a shallow, spatially large meson the two displayed terms
scale identically as the inverse size squared; their ratio is $g_K$ in
Eq.~\eqref{eq:central}. We use the decay and matching conditions of
Appendix~\ref{app:scaling} for the asymptotic laws below.
{An independent microscopic calculation through fourth order,
including separation-dependent hopping, gives
$A^{[4]}=2Jm_z^{2,[4]}/3$
(Appendix~\ref{app:weakfieldO4}).
A finite-ring estimator that cancels the threshold offset tests
this relation directly in the effective kernel
(Appendix~\ref{app:tail-difference}).}

\begin{figure*}[!t]
\centering
\includegraphics{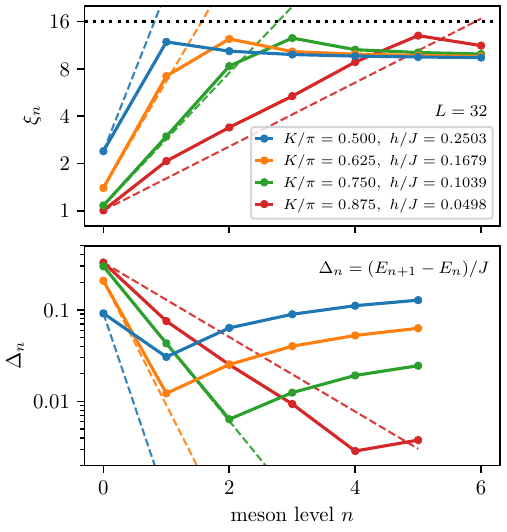}\hfill
\includegraphics{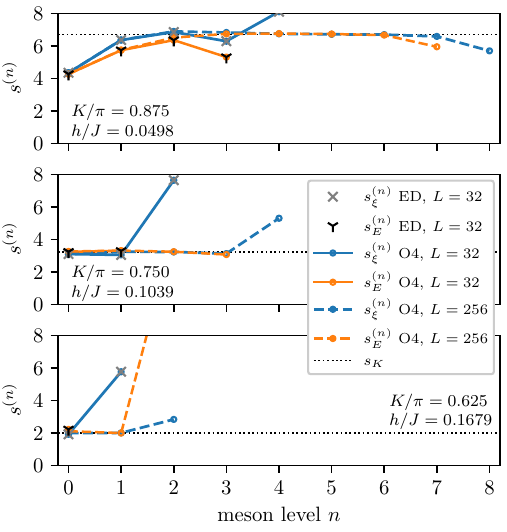}
\caption{Energy and size tests near the fitted $C_K=0$ contour.
Left: conditional rms separations $\xi_n$ and consecutive spacings
$\Delta_n=(E_{n+1}-E_n)/J$ from full-spin ED at $L=32$ for the four
pairs specified in Appendix~\ref{app:curvature_tuned}. Dashed guides show the predicted geometric scaling; the dotted line marks the exact bound
$L/2$. Right: effective size and spacing exponents
(Appendix~\ref{app:ED}) for the three larger momenta. Full-spin ED at
$L=32$ is compared with the fourth-order (O4) effective Hamiltonian at
$L=32$ and with its $L=256$ extension; dotted lines give the independent
$s_K$. The common intermediate plateau lies between core-dominated deep
levels and finite-ring bending of the shallowest levels.
Each dashed guide is normalized at its first displayed intermediate
level; its slope is fixed by the independently determined $s_K$ and is not
fit to the remaining meson data.}
\label{fig:DSI}
\end{figure*}

\emph{Scale anomaly and geometric mesons.---}
Write a bound-state energy as $E_a(K)=E_c(K)-\delta_a(K)$, with
$\delta_a=D_K\kappa_a^2>0$, and order states toward threshold as $a$
increases. At zero binding, the inverse-square equation gives
$f(x)\propto x^{1/2\pm\sqrt{1/4-g_K}}$.
For $g_K>1/4$ the exponents are complex and a real solution oscillates as
$\sqrt{x}\sin[s_K\log(x/x_0)]$.
Each additional node requires the same interval $\pi/s_K$ in $\log x$.
Thus the boundary phase selects the discrete dilation
$\Lambda_K=\ee^{\pi/s_K}$
\cite{Case1950,EssinGriffiths2006,Beane2001,BawinCoon2003}.
The lattice core regularizes short distances; the hierarchy accumulates
at the dissociation threshold [Fig.~\ref{fig:model_schematic}(b)].
For $g_K\leq1/4$, this oscillatory mechanism is absent.

At finite binding the decaying wave function is
$\sqrt{x}K_{\ii s_K}(\kappa_a x)$, where $K_\nu$ is the modified Bessel
function. Matching it to the core fixes an overall energy scale while
preserving the node spacing in logarithmic distance
\cite{MuellerHo2004,HammerSwingle2006}.
Consequently, the binding energies and rms sizes
$\xi_a=\sqrt{\langle x^2\rangle_a}$ obey
\begin{equation}
\frac{\delta_{a+1}}{\delta_a}\longrightarrow\Lambda_K^{-2},
\qquad
\frac{\xi_{a+1}}{\xi_a}\longrightarrow\Lambda_K,
\qquad \Lambda_K=\ee^{\pi/s_K}.
\label{eq:ratios}
\end{equation}
The reciprocal energy and size factors are a direct signature of the same
logarithmic wave function. The solution also fixes the amplitude product
$\delta_a\xi_a^2\to2D_K(1+s_K^2)/3$
(Appendix~\ref{app:scaling}).

There is one level per interval $\pi/s_K$ in logarithmic size.
A finite ring therefore accommodates, per resolved threshold channel,
\begin{equation}
\cN_K(L)=\frac{s_K}{\pi}\log L+\cO(1),
\label{eq:count}
\end{equation}
consistent with the discrete inverse-square counting law~\cite{DamanikTeschl2007}.
Here $K$ and $h$ are fixed as $L$ grows; the additive constant contains
core and boundary details. Equations~\eqref{eq:ratios} and
\eqref{eq:count} therefore predict the same $s_K$ from three independent
threshold observables, while Eq.~\eqref{eq:central} determines it
microscopically.

{For the numerical test, we select $(K,h)$ near $\smash{C_K(h)=0}$ with
$D_K>0$, suppressing the leading nonquadratic kinetic correction.
These points are chosen from the one-kink threshold scan before
analyzing the meson spectra (Appendix~\ref{app:curvature_tuned}).
We use consecutive spacings $\Delta_n=(E_{n+1}-E_n)/J$ to avoid
estimating the continuum threshold from the meson levels.
In full-spin ED, $\xi_n$ is the rms value of $\min(x,L-x)$,
conditioned on configurations with exactly two walls.
The effective exponents
\begin{equation}
s_\xi^{(n)}=\frac{\pi}{\log(\xi_{n+1}/\xi_n)},\qquad
s_E^{(n)}=\frac{2\pi}{\log(\Delta_n/\Delta_{n+1})}
\end{equation}
should both approach the independently determined $s_K$.}

{Figure~\ref{fig:DSI} tests this common prediction. Deep levels
retain core sensitivity, while the shallowest states resolve the
ring, leaving an intermediate scaling window.
Along the selected contour, increasing $K$ reduces $D_K$ and
densifies the hierarchy: $\Lambda_K$ decreases from approximately
$10.4$ at $K=\pi/2$, where no intermediate window is resolved at
$L=32$, to $1.60$ at $K=7\pi/8$.
The fourth-order effective Hamiltonian reproduces the $L=32$
ED trends and finite-size turnover. Its extension to $L=256$
reveals a longer common plateau, most clearly at $K=3\pi/4$
and $7\pi/8$. The effective observables are transformed consistently
with the Hamiltonian; their construction and validation are given
in Appendix~\ref{app:weakfieldO4}.
The large two-wall weights of the ED states support their mesonic
character (Appendix~\ref{app:ED}).}

\emph{Momentum-driven onset.---}
If an isolated quadratic channel satisfies
$D_{K_\star}=4A=8Jm_z^2/3$ and $D'_{K_\star}\neq0$,
then, at fixed $h$, the distance $\Delta K>0$ into the supercritical regime gives
$s_K\simeq\sqrt{B_\star\Delta K}$,
where
$B_\star=|D'_{K_\star}|/(4D_{K_\star})$.

For the asymptotic ratios
$\rho_E(K)=\lim_{a\to\infty}\delta_{a+1}/\delta_a$ and
$\rho_\xi(K)=\lim_{a\to\infty}\xi_{a+1}/\xi_a$, this gives
\begin{equation}
\log\rho_E(K)\sim-\frac{2\pi}{\sqrt{B_\star\Delta K}},
\qquad
\log\rho_\xi(K)\sim\frac{\pi}{\sqrt{B_\star\Delta K}}.
\label{eq:essential}
\end{equation}
Thus the universal ratios acquire an essential onset: the first
supercritical levels are exponentially separated, while the counting
slope vanishes as $\sqrt{\Delta K}$. Absolute branch locations retain
the nonuniversal core matching phase (Appendix~\ref{app:scaling}).

The weak-field projection provides a controlled supercritical benchmark.
A boundary spin flip moves either kink by one site, giving
$D_K=2h\cos(K/2)+\cO(h^2/J)$ and
$m_z=1+\cO[(h/J)^2]$ \cite{Liu2019}; hence, at fixed $|K|<\pi$ and
small $h/J$, $g_K=J/[3h\cos(K/2)]+\cO(1)>1/4$. A finite-momentum onset
is therefore a prediction for a dressed-curvature crossing beyond this
leading limit. The $C_K=0$ contour used in Fig.~\ref{fig:DSI} serves a
different purpose: it suppresses kinetic corrections without tuning the
critical condition $A/D_K=1/4$.

\emph{{Threshold scattering}.---}
Above threshold, the relative wave functions are combinations of
$\sqrt{x}J_{\pm\ii s_K}(qx)$, where $q>0$ is the relative momentum.
The same core phase that quantizes bound states gives the asymptotic
scattering relation
\begin{equation}
S_K(\Lambda_K q)-S_K(q)\longrightarrow0
\quad(q\to0^+),\quad \Lambda_K=\ee^{\pi/s_K}.
\label{eq:scattering}
\end{equation}
{The relation is exact for the ideal inverse-square problem
\cite{HammerSwingle2006}; lattice and core corrections vanish
at threshold. Thus scattering provides an independent measurement
of the exponent governing the bound-state hierarchy
(Appendix~\ref{app:scaling}).}

\emph{Experimental prospects.---} Programmable platforms provide possible routes toward the marginal
profile. Multitone spin-dependent forces in trapped ions can in principle
synthesize prescribed coupling matrices \cite{Korenblit_2012,Yang:2020}, arbitrary
Ising connectivities have been demonstrated in small registers \cite{Lu2025},
and domain-wall dynamics and bound-state energies are already measurable
\cite{Tan2021}. Meson spectroscopy has also been demonstrated in Rydberg
atom arrays~\cite{Vovrosh2025Spectroscopy}. Pairwise-tunable interactions in atoms coupled to
photonic-crystal waveguides provide a complementary architecture
\cite{Hung2016}. Exact marginality is not required over a finite window:
for $\alpha=4+\eta$, the effective inverse-square coupling drifts only
logarithmically, so approximate geometric scaling persists while
$|\eta|\log(x/x_{\rm core})\ll1$, with a parametric number of visible
cycles $\cN_{\rm DSI}=\cO(s_K/|\eta|)$ before finite-size and kinetic
cutoffs enter (Appendix~\ref{app:scaling}).

A direct test combines momentum-resolved spectroscopy with
wall-separation readout. A weak transverse modulation
$\delta h_j(t)=\epsilon\cos(Kj-\omega t)$ probes $S^{xx}(K,\omega)$~\cite{KnauteHauke2022}; consecutive resonances in one symmetry sector give
the spacing ratios, while frequency-selective preparation followed by
$z$-basis imaging gives the conditional separation distribution and hence
the size ratios. The target $s_K$ is obtained independently from the
correlation plateau and the continuum curvature
$D_K=2\partial_K^2E_c(K)$ when equal kink momenta minimize the threshold.
Changing~$L$ provides the logarithmic counting test, and threshold
scattering provides the fourth observable.

\emph{Conclusion and outlook.---}
We have identified a microscopic connection between magnetic order and
discrete scale invariance. At the marginal exponent $\alpha=4$, the
internal motion of a long-range Ising meson supports the supercritical
inverse-square scale anomaly, with an exponent fixed by the spontaneous
magnetization and dressed threshold curvature. This single exponent
connects geometric binding energies and sizes, logarithmic level
accumulation, and log-periodic threshold scattering. The independent
energy--size comparison supports this common scaling in the accessible
intermediate window, while counting and threshold scattering
provide further parameter-free tests. The construction identifies discrete scale
invariance as an emergent property of a two-kink magnetic composite rather
than an externally tuned few-body resonance.

Meson dynamics following Kibble--Zurek defect production has recently
been studied in a longitudinally biased Ising chain on a quantum
annealer~\cite{Bayocboc2026PostCritical}. The geometric hierarchy raises
a further question: how do domain walls
produced during a quantum critical ramp populate geometrically
spaced meson states? The Kibble--Zurek length~\cite{Jaschke2017}
could compete with their internal sizes, potentially producing
log-periodic dependence of meson populations or domain-wall
correlations on ramp time. In equilibrium, the corresponding
question is how bound levels and scattering states jointly enter
dilute-kink thermodynamics. These directions connect universal
defect production to the internal scaling and collective behavior of
the composites that defects form.

\emph{Acknowledgments.---} A.A.S. was funded by the Deutsche Forschungsgemeinschaft (DFG, German Research Foundation) under Project No.~557852701. D.V.V. acknowledges support from the Quantum Science
and Technology-National Science and Technology Major Project (Grant No. 2021ZD0301900).

\emph{Data availability.---} The numerical data and custom analysis code
supporting the findings of this Letter are available from the authors upon
reasonable request.

\bibliography{mesons}

@article{Jaschke2017,
	author = {Jaschke, Daniel and Maeda, Kenji and Whalen, Joseph D and Wall, Michael L and Carr, Lincoln D},
	doi = {10.1088/1367-2630/aa65bc},
	journal = {New Journal of Physics},
	month = {mar},
	number = {3},
	pages = {033032},
	publisher = {IOP Publishing},
	title = {Critical phenomena and Kibble--Zurek scaling in the long-range quantum Ising chain},
	url = {https://doi.org/10.1088/1367-2630/aa65bc},
	volume = {19},
	year = {2017}}

@article{McCoyWu1978,
	author = {McCoy, B. M. and Wu, T. T.},
	doi = {10.1103/PhysRevD.18.1259},
	journal = {Phys. Rev. D},
	pages = {1259--1267},
	title = {Two-dimensional {Ising} field theory in a magnetic field: Breakup of the cut in the two-point function},
	url = {https://doi.org/10.1103/PhysRevD.18.1259},
	volume = {18},
	year = {1978}}

@article{Zamolodchikov1989,
	author = {Zamolodchikov, A. B.},
	doi = {10.1142/S0217751X8900176X},
	journal = {Int. J. Mod. Phys. A},
	pages = {4235--4248},
	title = {Integrals of motion and {$S$}-matrix of the (scaled) {$T=T_c$} {Ising} model with magnetic field},
	url = {https://doi.org/10.1142/S0217751X8900176X},
	volume = {4},
	year = {1989}}

@article{Rutkevich2005,
	author = {Rutkevich, S. B.},
	doi = {10.1103/PhysRevLett.95.250601},
	journal = {Phys. Rev. Lett.},
	pages = {250601},
	title = {Large-{$n$} excitations in the ferromagnetic {Ising} field theory in a weak magnetic field: Mass spectrum and decay widths},
	url = {https://doi.org/10.1103/PhysRevLett.95.250601},
	volume = {95},
	year = {2005}}

@article{Rutkevich2008,
	author = {Rutkevich, S. B.},
	doi = {10.1007/s10955-008-9495-1},
	journal = {J. Stat. Phys.},
	pages = {917--939},
	title = {Energy spectrum of bound spinons in the quantum {Ising} spin-chain ferromagnet},
	url = {https://doi.org/10.1007/s10955-008-9495-1},
	volume = {131},
	year = {2008}}

@article{Coldea2010,
	author = {Coldea, R. and Tennant, D. A. and Wheeler, E. M. and Wawrzynska, E. and Prabhakaran, D. and Telling, M. and Habicht, K. and Smeibidl, P. and Kiefer, K.},
	doi = {10.1126/science.1180085},
	journal = {Science},
	pages = {177--180},
	title = {Quantum criticality in an {Ising} chain: Experimental evidence for emergent {$E_8$} symmetry},
	url = {https://doi.org/10.1126/science.1180085},
	volume = {327},
	year = {2010}}

@article{James2019,
	author = {James, A. J. A. and Konik, R. M. and Robinson, N. J.},
	doi = {10.1103/PhysRevLett.122.130603},
	journal = {Phys. Rev. Lett.},
	pages = {130603},
	title = {Nonthermal states arising from confinement in one and two dimensions},
	url = {https://doi.org/10.1103/PhysRevLett.122.130603},
	volume = {122},
	year = {2019}}

@article{Robinson2019,
	author = {Robinson, N. J. and James, A. J. A. and Konik, R. M.},
	doi = {10.1103/PhysRevB.99.195108},
	journal = {Phys. Rev. B},
	pages = {195108},
	title = {Signatures of rare states and thermalization in a theory with confinement},
	url = {https://doi.org/10.1103/PhysRevB.99.195108},
	volume = {99},
	year = {2019}}

@article{Verdel2020,
	author = {Verdel, R. and Liu, F. and Whitsitt, S. and Gorshkov, A. V. and Heyl, M.},
	doi = {10.1103/PhysRevB.102.014308},
	journal = {Phys. Rev. B},
	pages = {014308},
	title = {Real-time dynamics of string breaking in quantum spin chains},
	url = {https://doi.org/10.1103/PhysRevB.102.014308},
	volume = {102},
	year = {2020}}

@article{Karpov2022,
	author = {Karpov, P. I. and Zhu, G.-Y. and Heller, M. P. and Heyl, M.},
	doi = {10.1103/PhysRevResearch.4.L032001},
	journal = {Phys. Rev. Research},
	pages = {L032001},
	title = {Spatiotemporal dynamics of particle collisions in quantum spin chains},
	url = {https://doi.org/10.1103/PhysRevResearch.4.L032001},
	volume = {4},
	year = {2022}}

@article{Liu2019,
	author = {Liu, F. and Lundgren, R. and Titum, P. and Pagano, G. and Zhang, J. and Monroe, C. and Gorshkov, A. V.},
	doi = {10.1103/PhysRevLett.122.150601},
	journal = {Phys. Rev. Lett.},
	pages = {150601},
	title = {Confined quasiparticle dynamics in long-range interacting quantum spin chains},
	url = {https://doi.org/10.1103/PhysRevLett.122.150601},
	volume = {122},
	year = {2019}}

@article{Defenu2019,
	author = {Defenu, N. and Enss, T. and Halimeh, J. C.},
	doi = {10.1103/PhysRevB.100.014434},
	journal = {Phys. Rev. B},
	pages = {014434},
	title = {Dynamical criticality and domain-wall coupling in long-range {Hamiltonians}},
	url = {https://doi.org/10.1103/PhysRevB.100.014434},
	volume = {100},
	year = {2019}}

@article{Vovrosh2022,
	author = {Vovrosh, J. and Mukherjee, R. and Bastianello, A. and Knolle, J.},
	doi = {10.1103/PRXQuantum.3.040309},
	journal = {PRX Quantum},
	pages = {040309},
	title = {Dynamical hadron formation in long-range interacting quantum spin chains},
	url = {https://doi.org/10.1103/PRXQuantum.3.040309},
	volume = {3},
	year = {2022}}

@article{Case1950,
	author = {Case, K. M.},
	doi = {10.1103/PhysRev.80.797},
	journal = {Phys. Rev.},
	pages = {797--806},
	title = {Singular potentials},
	url = {https://doi.org/10.1103/PhysRev.80.797},
	volume = {80},
	year = {1950}}

@article{Frank1971,
	author = {Frank, W. M. and Land, D. J. and Spector, R. M.},
	doi = {10.1103/RevModPhys.43.36},
	journal = {Rev. Mod. Phys.},
	pages = {36--98},
	title = {Singular potentials},
	url = {https://doi.org/10.1103/RevModPhys.43.36},
	volume = {43},
	year = {1971}}

@article{EssinGriffiths2006,
	author = {Essin, A. M. and Griffiths, D. J.},
	doi = {10.1119/1.2165248},
	journal = {Am. J. Phys.},
	pages = {109--117},
	title = {Quantum mechanics of the {$1/x^2$} potential},
	url = {https://doi.org/10.1119/1.2165248},
	volume = {74},
	year = {2006}}

@article{Camblong2000,
	author = {Camblong, H. E. and Epele, L. N. and Fanchiotti, H. and {Garc\'ia Canal}, C. A.},
	doi = {10.1103/PhysRevLett.85.1590},
	journal = {Phys. Rev. Lett.},
	pages = {1590},
	title = {Renormalization of the inverse-square potential},
	url = {https://doi.org/10.1103/PhysRevLett.85.1590},
	volume = {85},
	year = {2000}}

@article{Beane2001,
	author = {Beane, S. R. and Bedaque, P. F. and Childress, L. and Kryjevski, A. and McGuire, J. and van Kolck, U.},
	doi = {10.1103/PhysRevA.64.042103},
	journal = {Phys. Rev. A},
	pages = {042103},
	title = {Singular potentials and limit cycles},
	url = {https://doi.org/10.1103/PhysRevA.64.042103},
	volume = {64},
	year = {2001}}

@article{BawinCoon2003,
	author = {Bawin, M. and Coon, S. A.},
	doi = {10.1103/PhysRevA.67.042712},
	journal = {Phys. Rev. A},
	pages = {042712},
	title = {Singular inverse square potential, limit cycles, and self-adjoint extensions},
	url = {https://doi.org/10.1103/PhysRevA.67.042712},
	volume = {67},
	year = {2003}}

@misc{MuellerHo2004,
	archiveprefix = {arXiv},
	author = {Mueller, E. J. and Ho, T.-L.},
	eprint = {cond-mat/0403283},
	title = {Renormalization-group limit cycles in quantum-mechanical problems},
	url = {https://arxiv.org/abs/cond-mat/0403283},
	year = {2004}}

@article{HammerSwingle2006,
	author = {Hammer, H.-W. and Swingle, B. G.},
	doi = {10.1016/j.aop.2005.04.017},
	journal = {Ann. Phys.},
	pages = {306--317},
	title = {On the limit cycle for the {$1/r^2$} potential in momentum space},
	url = {https://doi.org/10.1016/j.aop.2005.04.017},
	volume = {321},
	year = {2006}}

@article{DamanikTeschl2007,
	author = {Damanik, D. and Teschl, G.},
	doi = {10.1090/S0002-9939-06-08550-9},
	journal = {Proc. Amer. Math. Soc.},
	pages = {1123--1127},
	title = {Bound states of discrete {Schr\"odinger} operators with super-critical inverse square potentials},
	url = {https://doi.org/10.1090/S0002-9939-06-08550-9},
	volume = {135},
	year = {2007}}

@article{MorozSchmidt2010,
	author = {Moroz, S. and Schmidt, R.},
	doi = {10.1016/j.aop.2009.10.002},
	journal = {Ann. Phys.},
	pages = {491--513},
	title = {Nonrelativistic inverse square potential, scale anomaly, and complex extension},
	url = {https://doi.org/10.1016/j.aop.2009.10.002},
	volume = {325},
	year = {2010}}

@article{Efimov1970,
	author = {Efimov, V.},
	doi = {10.1016/0370-2693(70)90349-7},
	journal = {Phys. Lett. B},
	pages = {563--564},
	title = {Energy levels arising from resonant two-body forces in a three-body system},
	url = {https://doi.org/10.1016/0370-2693(70)90349-7},
	volume = {33},
	year = {1970}}

@article{BraatenHammer2006,
	author = {Braaten, E. and Hammer, H.-W.},
	doi = {10.1016/j.physrep.2006.03.001},
	journal = {Phys. Rep.},
	pages = {259--390},
	title = {Universality in few-body systems with large scattering length},
	url = {https://doi.org/10.1016/j.physrep.2006.03.001},
	volume = {428},
	year = {2006}}

@article{SunFengZhang2026,
	author = {Sun, N. and Feng, L. and Zhang, P.},
	doi = {10.1038/s42005-026-02580-0},
	journal = {Commun. Phys.},
	pages = {146},
	title = {{Efimov} effect in long-range quantum spin chains},
	url = {https://doi.org/10.1038/s42005-026-02580-0},
	volume = {9},
	year = {2026}}

@article{SunFengZhangImpurity2026,
	author = {Sun, N. and Feng, L. and Zhang, P.},
	doi = {10.1103/c7lm-323v},
	journal = {Phys. Rev. B},
	pages = {134442},
	title = {Universal bound states in long-range spin chains with an impurity},
	url = {https://doi.org/10.1103/c7lm-323v},
	volume = {113},
	year = {2026}}

@article{Vanderstraeten2018,
	author = {Vanderstraeten, L. and {Van Damme}, M. and B{\"u}chler, H. P. and Verstraete, F.},
	doi = {10.1103/PhysRevLett.121.090603},
	journal = {Phys. Rev. Lett.},
	pages = {090603},
	title = {Quasiparticles in quantum spin chains with long-range interactions},
	url = {https://doi.org/10.1103/PhysRevLett.121.090603},
	volume = {121},
	year = {2018}}

@article{DefenuRMP2023,
	author = {Defenu, N. and Donner, T. and Macr{\`\i}, T. and Pagano, G. and Ruffo, S. and Trombettoni, A.},
	doi = {10.1103/RevModPhys.95.035002},
	journal = {Rev. Mod. Phys.},
	pages = {035002},
	title = {Long-range interacting quantum systems},
	url = {https://doi.org/10.1103/RevModPhys.95.035002},
	volume = {95},
	year = {2023}}

@article{Feshbach1958,
	author = {Feshbach, H.},
	doi = {10.1016/0003-4916(58)90007-1},
	journal = {Ann. Phys.},
	pages = {357--390},
	title = {Unified theory of nuclear reactions},
	url = {https://doi.org/10.1016/0003-4916(58)90007-1},
	volume = {5},
	year = {1958}}

@article{Feshbach1962,
	author = {Feshbach, H.},
	doi = {10.1016/0003-4916(62)90221-X},
	journal = {Ann. Phys.},
	pages = {287--313},
	title = {A unified theory of nuclear reactions. {II}},
	url = {https://doi.org/10.1016/0003-4916(62)90221-X},
	volume = {19},
	year = {1962}}

@article{KnauteHauke2022,
	author = {Knaute, J. and Hauke, P.},
	doi = {10.1103/PhysRevA.105.022616},
	journal = {Phys. Rev. A},
	pages = {022616},
	title = {Relativistic meson spectra on ion-trap quantum simulators},
	url = {https://doi.org/10.1103/PhysRevA.105.022616},
	volume = {105},
	year = {2022}}

@article{Weinberg2017QuSpin,
	author = {Weinberg, P. and Bukov, M.},
	doi = {10.21468/SciPostPhys.2.1.003},
	journal = {SciPost Phys.},
	pages = {003},
	title = {{QuSpin}: a {Python} package for dynamics and exact diagonalisation of quantum many body systems part {I}: spin chains},
	url = {https://doi.org/10.21468/SciPostPhys.2.1.003},
	volume = {2},
	year = {2017}}

@article{Bravyi2011SW,
	author = {Bravyi, S. and DiVincenzo, D. P. and Loss, D.},
	doi = {10.1016/j.aop.2011.06.004},
	journal = {Ann. Phys.},
	pages = {2793--2826},
	title = {{Schrieffer--Wolff} transformation for quantum many-body systems},
	url = {https://doi.org/10.1016/j.aop.2011.06.004},
	volume = {326},
	year = {2011}}

@article{Hung2016,
	author = {Hung, C.-L. and Gonz{\'a}lez-Tudela, Alejandro and Cirac, J. Ignacio and Kimble, H. J.},
	doi = {10.1073/pnas.1603777113},
	journal = {Proc. Natl. Acad. Sci. U.S.A.},
	number = {34},
	pages = {E4946--E4955},
	title = {Quantum spin dynamics with pairwise-tunable, long-range interactions},
	url = {https://doi.org/10.1073/pnas.1603777113},
	volume = {113},
	year = {2016}}

@article{Tan2021,
	author = {Tan, W. L. and Becker, P. and Liu, F. and Pagano, G. and Collins, K. S. and De, A. and Feng, L. and Kaplan, H. B. and Kyprianidis, A. and Lundgren, R. and Morong, W. and Whitsitt, S. and Gorshkov, A. V. and Monroe, C.},
	doi = {10.1038/s41567-021-01194-3},
	journal = {Nat. Phys.},
	number = {6},
	pages = {742--747},
	title = {Domain-wall confinement and dynamics in a quantum simulator},
	url = {https://doi.org/10.1038/s41567-021-01194-3},
	volume = {17},
	year = {2021}}

@article{Lu2025,
	author = {Lu, Yao and Chen, Wentao and Zhang, Shuaining and Zhang, Kuan and Zhang, Jialiang and Zhang, Jing-Ning and Kim, Kihwan},
	doi = {10.1103/PhysRevLett.134.050602},
	journal = {Phys. Rev. Lett.},
	pages = {050602},
	title = {Implementing Arbitrary {Ising} Models with a Trapped-Ion Quantum Processor},
	url = {https://doi.org/10.1103/PhysRevLett.134.050602},
	volume = {134},
	year = {2025}}

@article{Korenblit_2012,
	author = {Korenblit, S and Kafri, D and Campbell, W C and Islam, R and Edwards, E E and Gong, Z-X and Lin, G-D and Duan, L-M and Kim, J and Kim, K and Monroe, C},
	doi = {10.1088/1367-2630/14/9/095024},
	journal = {New J. Phys.},
	number = {9},
	pages = {095024},
	title = {Quantum simulation of spin models on an arbitrary lattice with trapped ions},
	url = {https://doi.org/10.1088/1367-2630/14/9/095024},
	volume = {14},
	year = {2012}}

@misc{FonsecaZamolodchikov2006,
	archiveprefix = {arXiv},
	author = {Fonseca, Pedro and Zamolodchikov, Alexander},
	eprint = {hep-th/0612304},
	title = {{Ising} spectroscopy {I}: Mesons at {$T<T_c$}},
	url = {https://arxiv.org/abs/hep-th/0612304},
	year = {2006}}

@article{Nishida2013,
	author = {Nishida, Yusuke and Kato, Yasuyuki and Batista, Cristian D.},
	doi = {10.1038/nphys2523},
	journal = {Nat. Phys.},
	pages = {93--97},
	title = {{Efimov} effect in quantum magnets},
	url = {https://doi.org/10.1038/nphys2523},
	volume = {9},
	year = {2013}}

@article{Ovdat2017,
	author = {Ovdat, O. and Mao, Jinhai and Jiang, Yuhang and Andrei, E. Y. and Akkermans, E.},
	doi = {10.1038/s41467-017-00591-8},
	journal = {Nat. Commun.},
	pages = {507},
	title = {Observing a scale anomaly and a universal quantum phase transition in graphene},
	url = {https://doi.org/10.1038/s41467-017-00591-8},
	volume = {8},
	year = {2017}}

@misc{SupplementalMaterial,
	note = {See Supplemental Material for the microscopic tail, finite-ring boundary conditions, dressed thresholds and size observables, fourth-order weak-field construction and validation, and derivations of the scaling laws.}}

@misc{Li2026QuantumCriticality,
	archiveprefix = {arXiv},
	author = {Zhiyi Li and Zhijie Fan and Kun Chen and Youjin Deng},
	eprint = {2606.22407},
	primaryclass = {cond-mat.stat-mech},
	title = {Perturbative Renormalization and Universality Diagram for Long-Range Quantum Criticality},
	url = {https://arxiv.org/abs/2606.22407},
	year = {2026}}

@article{GonzalezCuadra2025String,
	author = {Gonz{\'a}lez-Cuadra, Daniel and Hamdan, Majd and Zache, Torsten V. and Braverman, Boris and Kornja{\v c}a, Milan and Lukin, Alexander and Cant{\'u}, Sergio H. and Liu, Fangli and Wang, Sheng-Tao and Keesling, Alexander and Lukin, Mikhail D. and Zoller, Peter and Bylinskii, Alexei},
	doi = {10.1038/s41586-025-09051-6},
	journal = {Nature},
	pages = {321--326},
	title = {Observation of string breaking on a {(2 + 1)D} {Rydberg} quantum simulator},
	url = {https://doi.org/10.1038/s41586-025-09051-6},
	volume = {642},
	year = {2025}}

@misc{De2024String,
	archiveprefix = {arXiv},
	author = {De, Arinjoy and Lerose, Alessio and Luo, De and Surace, Federica M. and Schuckert, Alexander and Bennewitz, Elizabeth R. and Ware, Brayden and Morong, William and Collins, Kate S. and Davoudi, Zohreh and Gorshkov, Alexey V. and Katz, Or and Monroe, Christopher},
	eprint = {2410.13815},
	primaryclass = {quant-ph},
	title = {Observation of string-breaking dynamics in a quantum simulator},
	url = {https://arxiv.org/abs/2410.13815},
	year = {2024}}

@article{Bennewitz2025,
	author = {Bennewitz, Elizabeth R. and Ware, Brayden and Schuckert, Alexander and Lerose, Alessio and Surace, Federica M. and Belyansky, Ron and Morong, William and Luo, De and De, Arinjoy and Collins, Kate S. and Katz, Or and Monroe, Christopher and Davoudi, Zohreh and Gorshkov, Alexey V.},
	doi = {10.22331/q-2025-06-17-1773},
	journal = {Quantum},
	pages = {1773},
	title = {Simulating Meson Scattering on Spin Quantum Simulators},
	url = {https://doi.org/10.22331/q-2025-06-17-1773},
	volume = {9},
	year = {2025}}

@article{Surace2026,
	author = {Surace, Federica Maria and Lerose, Alessio and Katz, Or and Bennewitz, Elizabeth R. and Schuckert, Alexander and Luo, De and De, Arinjoy and Ware, Brayden and Morong, William and Collins, Kate and Monroe, Christopher and Davoudi, Zohreh and Gorshkov, Alexey V.},
	doi = {10.1103/c4zd-lbyq},
	journal = {PRX Quantum},
	pages = {020331},
	title = {String-Breaking Dynamics in Quantum Adiabatic and Diabatic Processes},
	url = {https://doi.org/10.1103/c4zd-lbyq},
	volume = {7},
	year = {2026}}

@misc{Vovrosh2025Spectroscopy,
	archiveprefix = {arXiv},
	author = {Vovrosh, Joseph and {de Hond}, Julius and Juli{\`a}-Farr{\'e}, Sergi and Knolle, Johannes and Dauphin, Alexandre},
	eprint = {2506.21299},
	note = {Accepted for publication in PRX Quantum},
	primaryclass = {quant-ph},
	title = {Meson spectroscopy of exotic symmetries of {Ising} criticality in {Rydberg} atom arrays},
	url = {https://arxiv.org/abs/2506.21299},
	year = {2025}}

@article{Kormos2017,
	author = {Kormos, M{\'a}rton and Collura, Mario and Tak{\'a}cs, G{\'a}bor and Calabrese, Pasquale},
	doi = {10.1038/nphys3934},
	journal = {Nat. Phys.},
	pages = {246--249},
	title = {Real-time confinement following a quantum quench to a non-integrable model},
	url = {https://doi.org/10.1038/nphys3934},
	volume = {13},
	year = {2017}}

@misc{Bayocboc2026PostCritical,
	archiveprefix = {arXiv},
	author = {Bayocboc, Jr., Francis A. and Dziarmaga, Jacek and Rams, Marek M. and Vodeb, Jaka},
	eprint = {2607.13842},
	primaryclass = {quant-ph},
	title = {Post-Critical Meson Dynamics of {Kibble--Zurek} Excitations in a 5,564-Qubit Quantum Annealer},
	url = {https://arxiv.org/abs/2607.13842},
	year = {2026}}

@article{Yang:2020,
	author = {Yang, Dayou and Grankin, Andrey and Sieberer, Lukas M. and Vasilyev, Denis V. and Zoller, Peter},
	date = {2020/02/07},
	doi = {10.1038/s41467-020-14489-5},
	id = {Yang2020},
	isbn = {2041-1723},
	journal = {Nature Communications},
	number = {1},
	pages = {775},
	title = {Quantum non-demolition measurement of a many-body Hamiltonian},
	url = {https://doi.org/10.1038/s41467-020-14489-5},
	volume = {11},
	year = {2020}}

@article{SuraceLerose2021,
	doi = {10.1088/1367-2630/abfc40},
	url = {https://doi.org/10.1088/1367-2630/abfc40},
	year = {2021},
	month = {jun},
	publisher = {IOP Publishing},
	volume = {23},
	number = {6},
	pages = {062001},
	author = {Surace, Federica Maria and Lerose, Alessio},
	title = {Scattering of mesons in quantum simulators},
	journal = {New Journal of Physics}
	}

@article{Lerose2020,
  title = {Quasilocalized dynamics from confinement of quantum excitations},
  author = {Lerose, Alessio and Surace, Federica M. and Mazza, Paolo P. and Perfetto, Gabriele and Collura, Mario and Gambassi, Andrea},
  journal = {Phys. Rev. B},
  volume = {102},
  issue = {4},
  pages = {041118(R)},
  numpages = {7},
  year = {2020},
  month = {Jul},
  publisher = {American Physical Society},
  doi = {10.1103/PhysRevB.102.041118},
  url = {https://link.aps.org/doi/10.1103/PhysRevB.102.041118}
	}

\onecolumngrid
\clearpage

\setcounter{figure}{0}
\renewcommand{\thefigure}{S\arabic{figure}}
\renewcommand{\theHfigure}{S\arabic{figure}}
\setcounter{table}{0}
\renewcommand{\thetable}{S\Roman{table}}
\renewcommand{\theHtable}{S\arabic{table}}
\begin{center}
\textbf{Supplemental Material for ``Discrete Scale Invariance of Ising Mesons''}
\end{center}

{This Supplemental Material provides the derivations and numerical
methods supporting the main text.
Appendix~\ref{app:tail} derives the interaction tail and its
correlation reconstruction.
Appendix~\ref{app:ED} describes the finite-ring calculations,
dressed thresholds, and size observables.
Appendices~\ref{app:weakfieldO4} and \ref{app:fourth_order} develop
the fourth-order effective Hamiltonian, verify the relation between
the dressed attraction and magnetization, and obtain the one-kink
dispersion.
Appendix~\ref{app:curvature_tuned} explains the parameter selection
and compares perturbation theory with exact diagonalization.
Appendix~\ref{app:tail-difference} tests the dressed tail without
estimating the threshold energy.
Appendix~\ref{app:scaling} derives the scaling, counting, and
scattering laws and discusses their corrections, the conditional
momentum-driven onset, and departures from marginality.}

\appendix
\setcounter{secnumdepth}{2}

\section{Microscopic interaction and correlation reconstruction}
\label{app:tail}
\label{app:inverse-square}

For an integer domain length $x$, there are $2\min(r,x)$ bonds of range $r$ crossing its two interfaces. Each changes its classical energy by $2J/r^\alpha$, giving
\begin{equation}
V_\alpha^{(0)}(x)=4J\left[\sum_{r=1}^{x}r^{1-\alpha}
+x\sum_{r=x+1}^{\infty}r^{-\alpha}\right].
\label{app:Vbare}
\end{equation}
For $\alpha>2$, the threshold is $V_\alpha^{(0)}(\infty)=4J\zeta(\alpha-1)$ and
\begin{equation}
V_\alpha^{(0)}(x)-V_\alpha^{(0)}(\infty)
=-\frac{4Jx^{2-\alpha}}{(\alpha-1)(\alpha-2)}+o(x^{2-\alpha}).
\label{app:generalalpha}
\end{equation}
At $\alpha=4$, the exact difference and its large-$x$ expansion are
\begin{align}
V_4^{(0)}(\infty)-V_4^{(0)}(x)
&=4J[\zeta(3,x+1)-x\zeta(4,x+1)],\label{app:Hurwitz}\\
V_4^{(0)}(x)-V_4^{(0)}(\infty)
&=-\frac{2J}{3x^2}+\frac{J}{3x^4}-\frac{J}{3x^6}
+\cO(x^{-8}).\label{app:baretail}
\end{align}
Thus $\alpha=4$ selects an inverse-square attraction, with a bare $x^{-4}$ correction.

For dressed interfaces, work in the ordered phase with localized cores
and the power-decaying corrections assumed in Eq.~(3).
After subtracting the two isolated-wall energies, corrections to the
leading bulk crossing-bond contribution are $\cO(x^{-2-\omega})$ for
some $\omega>0$ at $\alpha=4$, including connected correlations and
interface deformation. These are spatial assumptions on the channel;
a bulk gap alone does not establish them or imply exponential
correlation decay in this long-range model.

Far from a core, the two quantum vacua have magnetizations $\pm m_z$.
The spin product on a long bond therefore approaches $+m_z^2$ within
one vacuum and $-m_z^2$ across an interface. Its energy change is
$2Jm_z^2/r^\alpha$. Thus $m_z^2$ enters as a property of the dressed
vacua, and the crossing-bond count gives at $\alpha=4$
\begin{equation}
V_{\rm rel}(\infty)-V_{\rm rel}(x)
=\frac{2Jm_z^2}{3x^2}+\cO(x^{-2-\omega}).
\label{app:dressedtail}
\end{equation}
Long bonds joining bulk regions determine this leading term. Local
interface structure fixes the core energies and, under the stated
assumptions, contributes only subleading separation dependence after
isolated-wall subtraction. The argument is nonperturbative in $h$ within
this regime. Appendix~\ref{app:weakfieldO4} verifies the same relation
microscopically through fourth order, including the contributions beyond
the classical diagonal potential.

As a supporting construction, Fig.~2 reconstructs
the crossing-bond energy from ground-state correlations. In a
symmetry-resolved periodic ground state $|\Omega\rangle$, the one-point
magnetization vanishes; instead we use
\begin{equation}
C_{zz}(r)=\frac1L\sum_i\langle\Omega|\sigma_i^z\sigma_{i+r}^z|\Omega\rangle,
\quad m_z^2=\lim_{r\to\infty}\lim_{L\to\infty}C_{zz}(r).
\label{app:Czz}
\end{equation}
Reversing an interval by $U_x=\prod_{i=1}^{x}\sigma_i^x$ leaves the transverse-field term invariant and reverses precisely the crossing Ising bonds. On the infinite chain, its excess energy is
\begin{equation}
V_{\rm corr}(x)=4J\left[\sum_{r=1}^{x}\frac{C_{zz}(r)}{r^3}
+x\sum_{r=x+1}^{\infty}\frac{C_{zz}(r)}{r^4}\right].
\label{app:Vnumerical}
\end{equation}
This domain-flipped state can have a different core structure from the
relaxed dressed channel. The correlation plateau fixes its leading
inverse-square amplitude. The plateau part gives
\begin{equation}
\bigl[V_{\rm corr}(\infty)-V_{\rm corr}(x)\bigr]_{\rm plateau}
=\frac{2Jm_z^2}{3x^2}-\frac{Jm_z^2}{3x^4}+\cO(x^{-6}).
\label{app:plateau-lattice}
\end{equation}
The $x^{-4}$ coefficient belongs to this plateau contribution, not to a
universal next term of $V_{\rm rel}$.

For the displayed reconstruction, correlations are available to $r_{\max}=L/2$ and are continued by $\widehat C_{zz}(r)=m_z^2$ beyond that distance. For $x\leq r_{\max}$ it is convenient to evaluate the difference directly:
\begin{equation}
V_{\rm corr}(\infty)-V_{\rm corr}(x)=
4J\bigg[\sum_{r=x+1}^{r_{\max}}\frac{(r-x)C_{zz}(r)}{r^4}
+m_z^2\{\zeta(3,r_{\max}+1)
-x\zeta(4,r_{\max}+1)\}\bigg].
\label{app:inverse-square-test}
\end{equation}
The continuation supplies the asymptotic amplitude. The plot therefore
illustrates the finite-distance approach to the bond-counting tail,
including the approach from below of the plateau contribution in
Eq.~\eqref{app:plateau-lattice}. The effective-kernel test in
Appendix~\ref{app:tail-difference} and the meson energies and conditional
sizes in Fig.~3 test the dressed-channel account separately.

\section{Finite-ring ED, dressed thresholds, and size observables}
\label{app:ED}
\label{app:twisted_threshold}

\subsection{Boundary conditions and threshold extraction}

The periodic ED implementation uses the image-summed coupling
\begin{equation}
J_L^+(r)=J\sum_{m\in\mathbb Z}|r+mL|^{-\alpha}=\frac{J}{L^\alpha}\left[\zeta\!\left(\alpha,\frac rL\right)
+\zeta\!\left(\alpha,1-\frac rL\right)\right]
\label{app:periodic_coupling}
\end{equation}
for $1\leq r<L$. The antiperiodic continuation $s_{i+L}=-s_i$ instead assigns $(-1)^m$ to image $m$, giving $J_L^-(r)=J\sum_m(-1)^m|r+mL|^{-\alpha}$. These two conventions have the same infinite-chain interaction. Finite-ring perturbation theory must use the same couplings as the ED dataset being compared.

The twisted ring contains an odd number of walls. Its lowest isolated dispersive band describes a dressed kink. Translation through the seam acts as
\begin{equation}
\widetilde T|s_0,\ldots,s_{L-1}\rangle
=|s_1,\ldots,s_{L-1},-s_0\rangle,
\quad \widetilde T^L=\mathcal P,
\label{app:twisted_translation}
\end{equation}
where $\mathcal P=\prod_i\sigma_i^x$. Hence the single-kink momentum is
$p_\nu=\pi\nu/L$, with spin-flip eigenvalue $z=(-1)^\nu$.
The lowest eigenvalue $E_{\rm tw}(p_\nu)$ determines the one-kink
dispersion up to a momentum-independent reference energy. When used in
excitation energies below, $\varepsilon(p)$ is measured above the
periodic vacuum with the same bulk-energy convention. For two distant kinks,
\begin{equation}
E_2(K,q)=\varepsilon(K/2+q)+\varepsilon(K/2-q).
\label{app:two_kink_threshold}
\end{equation}
If $q=0$ is the isolated global minimum, its local expansion is
\begin{equation}
E_2(K,q)=E_c(K)+D_Kq^2+C_Kq^4+\cO(q^6),
\qquad D_K>0,
\label{eq:threshold}
\end{equation}
where $D_K=\varepsilon''(K/2)$,
$C_K=\varepsilon^{(4)}(K/2)/12$, and
$F_K=\varepsilon^{(6)}(K/2)/360$ is the sixth-order coefficient.
The equal-momentum threshold satisfies $E_c(K)=2\varepsilon(K/2)$,
so $D_K=2\partial_K^2E_c(K)$ when differentiation is performed at fixed
$h$. Positivity of the local curvature must be accompanied by a check
that no lower minimum occurs at another relative momentum.

For $K=2\pi k/L$, $p=K/2$ lies exactly on the twisted grid. With $d_p=\pi/L$, the three symmetric differences
\begin{align}
Y_j&=E_{\rm tw}(p+jd_p)+E_{\rm tw}(p-jd_p)-2E_{\rm tw}(p),\notag\\
Y_j&=D_K(jd_p)^2+C_K(jd_p)^4+F_K(jd_p)^6+\cO(d_p^8),
\label{app:twisted_fit}
\end{align}
for $j=1,2,3$ give a linear system for the three coefficients. The resulting values are finite-stencil estimates, not exact derivatives. Varying stencil width and $L$ tests their errors. Neither meson energies nor sizes enter this determination. A negative fitted $D_K$, or a change of the minimizing relative momentum, invalidates the $q=0$ reduction; such points in the broad scan do not define a physical anomaly exponent.

For the $L=32$ spectra, we used SciPy's
\texttt{scipy.sparse.linalg.eigsh} solver with the requested tolerance
$\texttt{tol}=10^{-12}$ in the relevant translation-momentum sectors
$K$, restricting to the even spin-flip sector $z=+1$; reflection
symmetry was not imposed at generic momentum. For the $L=36$
correlation reconstruction, we targeted the $K=0$, reflection parity $+1$, $z=+1$
ground-state sector and applied the same solver to a matrix-free
Hamiltonian represented by a
\texttt{scipy.sparse.linalg.LinearOperator}.

The channel reduction can also be expressed by the Feshbach operator
\begin{equation}
H_{\rm eff}(E)=PHP+PHQ(E-QHQ)^{-1}QHP.
\label{app:Feshbach}
\end{equation}
Here $P$ denotes the asymptotic dressed channel. If a finite energy
interval around $E_c(K)$ is separated from all eliminated continua in
the same symmetry sector, the analytic part renormalizes the threshold,
kinetic coefficients, and core matching condition. An open decay channel
would instead require a multichannel or resonance description.
Appendix~\ref{app:weakfieldO4} constructs an energy-independent weak-field
version using the bare two-wall subspace.

\subsection{Sizes, spacings, and the finite-size window}

In a periodic spin configuration with exactly two walls, define $X=\min(x,L-x)$. Let $P_2$ be the projector onto those configurations and extend $X$ by zero outside this subspace. The ED observables are
\begin{equation}
W_{2,n}=\langle\psi_n|P_2|\psi_n\rangle,
\qquad
\xi_n^2=\frac{\langle\psi_n|X^2|\psi_n\rangle}{W_{2,n}}.
\label{app:EDsize}
\end{equation}
The weight is the physical probability of exactly two walls in the
full-spin state.
Thus the plotted size is a conditional rms separation, with the exact
geometric bound $\xi_n\leq L/2$. The dotted lines in
Figs.~3 and \ref{fig:K0_meson_scaling} mark this upper bound;
an extended state's rms size can saturate below it. States are ordered
by energy within the same resolved symmetry sector and meson sequence.
The ratio estimators require consecutive members of that sequence.
Here $E_n$ denotes an absolute eigenvalue and
$e_n=E_n-E_{\rm vac}$ the corresponding excitation energy.
The level index $n$ starts at zero for the lowest member of the meson
sequence, rather than for the ground state of the periodic chain.

To avoid estimating the continuum threshold from finite-ring meson energies, we use
\begin{equation}
\Delta_n=\frac{E_{n+1}-E_n}{J},\qquad s_\xi^{(n)}=\frac{\pi}{\log(\xi_{n+1}/\xi_n)},\qquad
s_E^{(n)}=\frac{2\pi}{\log(\Delta_n/\Delta_{n+1})}.
\label{app:effective_exponents_tuned}
\end{equation}
If $\delta_n=E_c-E_n=\delta_\star\rho^n$ with
$\rho=\ee^{-2\pi/s_K}$, then
$\Delta_n=(1-\rho)\delta_n/J$ and both estimators tend to $s_K$.
Near saturation, $\log(\xi_{n+1}/\xi_n)$ tends to zero and the size
estimator becomes large. The energy estimator similarly loses its
threshold interpretation when spacings turn upward. These features
explain the departures in Fig.~3 and identify the useful
intermediate levels. The size estimator uses levels $n,n+1$; the
spacing estimator uses all three levels $n,n+1,n+2$.

\begin{figure}[!tbp]
\centering
\includegraphics[]{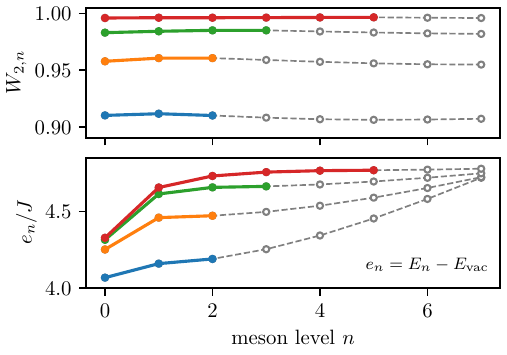}
\caption{Two-wall content and excitation energies of the full-spin ED
meson sequences at $L=32$, $\alpha=4$, and $z=+1$, with no reflection
restriction at these momenta. Top: physical probabilities $W_{2,n}$
[Eq.~\eqref{app:EDsize}]. Bottom: $e_n/J=(E_n-E_{\rm vac})/J$.
Blue, orange, green, and red correspond respectively to
$K/\pi=1/2,5/8,3/4,7/8$ and the fields in Table~\ref{tab:h0_curve}.
Filled colored markers specify levels retained for the scaling analysis in Fig.~3 in the main text.}
\label{fig:channel_isolation_check}
\end{figure}

Figure~\ref{fig:channel_isolation_check} gives the content and excitation
energies of the $L=32$ sequences in Fig.~3.
The large $W_{2,n}$ values support their two-wall character. As in
the threshold theory of the main text, the single-channel interpretation
additionally requires that no eliminated continuum be open in the same
$(K,z)$ sector; in the controlled weak-field regime this is the sector
continuously connected to the separated two-wall channel at $h=0$.
The marker-style transition in this figure is not used to define the
scaling window or to discard a level needed for an adjacent-spacing ratio.

\section{Canonical weak-field Hamiltonian and observables through fourth order}
\label{app:weakfieldO4}
\label{app:weakfieldO2}

We verify Appendix~\ref{app:tail}'s physical dressed-tail relation
microscopically through fourth order and construct the effective
dynamics and observables used in the quantitative tests. The vacuum
energy and its longitudinal-field derivative yield $m_z^{2,[4]}$;
a separate canonical two-wall calculation yields $A^{[4]}$. Their
coefficients are compared only after both calculations. The tail
calculation below is followed by a separate proof of its auxiliary bounds.

\subsection{Canonical construction through fourth order}

We use $u=h/J$ for numerical field values. For any formal series $Q$,
$[Q]_{\leq N}$ retains every power of $h$ through $h^N$ at fixed $J$
(equivalently every power of $u$ through $u^N$).
It is an exactly defined polynomial, distinct from its order-$N$
coefficient. Remainders below refer to untruncated quantities.

Write $H=H_0+hV$, with $V=-\sum_i\sigma_i^x$, and let $P$ now project onto all bare two-wall configurations. The state $|j,x\rangle$ reverses sites $j+1,\ldots,j+x$. Our momentum convention is
\begin{equation}
|K,x\rangle=\frac1{\sqrt L}\sum_j\ee^{-\ii K(j+x/2)}|j,x\rangle,
\qquad K=\frac{2\pi k}{L}.
\label{app:KxbasisO4}
\end{equation}
Boundary flips give $\langle K,x+1|hV|K,x\rangle=-2h\cos(K/2)$. With $f_0=0$ on the infinite half-line, the leading projection is
\begin{equation}
-2h\cos(K/2)(f_{x+1}+f_{x-1})+V_\alpha^{(0)}(x)f_x=Ef_x.
\label{app:discrete}
\end{equation}

For $Q=1-P$, choose the canonical anti-Hermitian generator $S=\sum_{n=1}^{4}h^nS_n$ with $PS_nP=QS_nQ=0$ and transform $\widetilde H=\ee^SH\ee^{-S}$. Define $H^{(0)}=H_0$, $H^{(1)}=V$, and $H^{(n>1)}=0$. Then
\begin{align}
\widetilde H_n=H^{(n)}+
\sum_{q=1}^{n}\frac1{q!}
\sum_{\substack{r_1,\ldots,r_q\geq1,\ s=0,1\\r_1+\cdots+r_q+s=n}}
\operatorname{ad}_{S_{r_1}}\cdots\operatorname{ad}_{S_{r_q}}H^{(s)},
\label{app:SWrecursionO4}
\end{align}
where $\operatorname{ad}_S O=[S,O]$. At order $n$, let $R_n$ be this expression with $S_n$ omitted. The off-block condition fixes
\begin{equation}
(S_n)_{ma}=\frac{(R_n)_{ma}}{E_m^{(0)}-E_a^{(0)}},
\qquad (S_n)_{am}=-(S_n)_{ma}^*,
\label{app:SrecursionO4}
\end{equation}
for $a\in P$, $m\in Q$. With $H_n=P\widetilde H_nP$, the retained Hamiltonian is
\begin{equation}
H_{\rm 2k}^{[N]}=PH_0P+\sum_{n=1}^{N}h^nH_n,
\qquad N=2,4.
\label{app:HeffO4}
\end{equation}
The expansion requires nonzero $P$--$Q$ denominators, with the field
small compared with the relevant gaps. Each denominator retains its
configuration-dependent energy. In particular,
\begin{equation}
(H_2)_{ab}=\frac12\sum_{m\in Q}V_{am}V_{mb}
\left[\frac1{E_a^{(0)}-E_m^{(0)}}+\frac1{E_b^{(0)}-E_m^{(0)}}\right].
\label{app:H2matrix}
\end{equation}
Equations~\eqref{app:SWrecursionO4}--\eqref{app:HeffO4} also specify $H_3$ and $H_4$, including paths that revisit $P$ and the associated normalization terms. Vacuum subtraction is carried out with a separate SW reduction retaining the two ferromagnetic configurations, through the same order.

\subsection{Path and momentum representation}

A four-flip path beginning and ending in $P$ reaches graph distance at most two from $P$. Hence the induced graph $\{s:d(s,P)\leq2\}$ suffices for all retained coefficients through fourth order. It contains the relevant zero-, four-, and six-wall virtual states, without requiring the full $2^L$ basis. For each order the exact momentum kernel is
\begin{equation}
(H_n)_{x'x}(K)=\sum_d\ee^{\ii K[d+(x'-x)/2]}
(H_n)_{(d,x'),(0,x)}.
\label{app:HnKO4}
\end{equation}
Away from short-domain and wrapping processes, second order contributes diagonal and $\Delta x=\pm2$ terms, third order contributes $\Delta x=\pm1,\pm3$, and fourth order contributes diagonal and $\Delta x=\pm2,\pm4$ terms. Their coefficients retain the separation dependence of the interaction. Annihilation into a vacuum at $x=1$ or $L-1$ and finite-ring coincidences are included by the same path rule; a far-field dispersion alone does not specify this core Hamiltonian.

Global Ising inversion acts as $\mathcal P|K,x\rangle=(-1)^k|K,L-x\rangle$. In parity sector $z=\pm1$, the folded basis is $[|K,x\rangle+z(-1)^k|K,L-x\rangle]/\sqrt2$ for $x<L/2$. At even $L$, the state $x=L/2$ is retained only if $z=(-1)^k$.

The finite matrix $H_{\rm 2k}^{[N]}-E_{\rm vac}^{[N]}$ is diagonalized without further expansion in the separation coordinate. The resulting meson spectrum therefore resums motion within the retained channel, while virtual excursions outside it are included only through order $h^N$. This distinction is essential for a geometric hierarchy that is nonanalytic in its microscopic coupling. A fourth-order Hamiltonian generically omits fifth-order matrix elements. Sixth-order errors for individual scalar observables require additional branch and symmetry conditions, as discussed in Sec.~\ref{app:quspin-validation}.

\subsection{Linked vacuum terms and magnetization}

On the infinite chain define the one- and two-spin-flip gaps
\begin{equation}
\Delta=4J\zeta(\alpha),\qquad \Delta_2(r)=2\Delta-4J/r^\alpha.
\end{equation}
These are classical vacuum excitation gaps, distinct from the
dimensionless level spacings $\Delta_n$. Flipping one spin reverses
both bonds at each range and costs $\Delta$. Adding two single-flip
costs counts the bond between the flipped spins twice, although it is
unchanged when both are flipped; subtracting $4J/r^\alpha$ gives
$\Delta_2(r)$.
The vacuum energy density through fourth order is
\begin{align}
e_{\rm vac}&=-J\zeta(\alpha)-\frac{h^2}{\Delta}
+\frac{h^4B}{\Delta^2}+\cO(h^6),\notag\\
B&=\frac1\Delta+2\sum_{r=1}^{\infty}
\left(\frac1\Delta-\frac2{\Delta_2(r)}\right).
\label{app:vacO4}
\end{align}
The repeated one-site walk gives $\Delta^{-3}$. For a pair at distance
$r$, the four irreducible walks contribute
$-4/[\Delta^2\Delta_2(r)]$, and the mixed normalization terms give
$2/\Delta^3$. There is one unordered pair per site at each positive
$r$, which yields Eq.~\eqref{app:vacO4}; the disconnected parts cancel
within each summand. Sec.~\ref{app:dtO4} derives the same terms
for unequal local gaps.

Introduce a longitudinal term $-\ell\sum_i\sigma_i^z$ to select the
positive vacuum, with the thermodynamic limit preceding $\ell\to0^+$.
The classical density becomes $-J\zeta(\alpha)-\ell$ and
$\Delta(\ell)=\Delta+2\ell$,
$\Delta_2(r,\ell)=\Delta_2(r)+4\ell$. Thus
$m_z=-\partial_\ell e_{\rm vac}|_{0^+}$; explicitly,
\begin{equation}
\partial_\ell\frac{B(\ell)}{\Delta(\ell)^2}\bigg|_{0^+}
=\frac{B'}{\Delta^2}-\frac{4B}{\Delta^3}.
\label{app:mz_derivative}
\end{equation}
Here $B'=\partial_\ell B|_{0^+}$, not a transverse-field derivative.
It follows that
\begin{align}
m_z&=1-\frac{2h^2}{\Delta^2}
+h^4\left(\frac{4B}{\Delta^3}-\frac{B'}{\Delta^2}\right)+\cO(h^6),\notag\\
B'&=-\frac2{\Delta^2}+2\sum_{r=1}^{\infty}
\left[-\frac2{\Delta^2}+\frac8{\Delta_2(r)^2}\right].
\label{app:mzO4}
\end{align}
The summands defining $B$ and $B'$ are $\cO(r^{-\alpha})$ and
the pair gaps are bounded away from zero for $\ell$ near $0^+$;
the differentiation therefore preserves absolute convergence.
Their units are $[B]=J^{-1}$ and $[B']=J^{-2}$.
Squaring the preceding series and discarding powers above four gives
\begin{equation}
m_z^{2,[4]}\equiv[m_z^2]_{\leq4}
=1-\frac{4h^2}{\Delta^2}
+h^4\left(\frac4{\Delta^4}
+\frac{8B}{\Delta^3}-\frac{2B'}{\Delta^2}\right).
\label{app:mz2B}
\end{equation}
In particular, the square of the quadratic term must be retained.
At $\alpha=4$, with $u=h/J$,
\begin{equation}
m_z^2=1-0.2134155430983u^2-0.0530246101889u^4+\cO(u^6).
\label{app:mz2O4}
\end{equation}
Evaluation of the sums gives these coefficients; the displayed remainder
belongs to the untruncated $m_z^2$, while Eq.~\eqref{app:mz2B} defines
the polynomial exactly.
The next
subsection derives the interaction coefficient from the two-wall
Hamiltonian and establishes its relation to the same $B$ and $B'$.

\subsection{Dressed interaction tail and comparison with magnetization}
\label{app:dtO4}

To verify the physical relation in Appendix~\ref{app:tail}, we calculate
the inverse-square attraction independently of the magnetization, using
the canonical Hamiltonian of
Eq.~\eqref{app:HeffO4} at $\alpha=4$ in the centered basis of
Eq.~\eqref{app:KxbasisO4}. Let $H_{\rm free,K}^{[4]}$ be the
two-independent-wall excitation kernel, including all hopping through
fourth order and with the common vacuum energy removed. Define
\begin{equation}
\mathcal W_K^{[4]}
=H_{\rm 2k,K}^{[4]}-E_{\rm vac}^{[4]}I-H_{\rm free,K}^{[4]}.
\label{dtO4:interaction}
\end{equation}
Its complete row sum determines the interaction coefficient. The free
row sum is $E_c^{[4]}(K)\equiv[E_c(K)]_{\leq4}$ at relative momentum
$q=0$, with the threshold interpretation subject to the stated minimum
conditions. Here $\mathcal W_K^{[4]}$ includes the inverse-square term;
$W_K$ denotes the continuum remainder and $W_{2,n}$ the physical
two-wall probability.

\paragraph{Connected background energies.}
Let $D_x$ have $s_i=-1$ on $1\leq i\leq x$ and $+1$ elsewhere.
The isolated left wall $\mathsf L$ has $s_i=+1$ for $i\leq0$ and
$-1$ for $i>0$; the right wall $\mathsf R$ has $s_i=-1$ for
$i\leq x$ and $+1$ for $i>x$. The fourth background $+$ is uniform.
For a background functional $F$, define
\begin{equation}
\mathcal I_xF\equiv F[D_x]-F[\mathsf L]-F[\mathsf R]+F[+].
\label{dtO4:connected-subtraction}
\end{equation}
The vacuum is added back because the two isolated-wall energies contain
it twice. A common bulk term and either additive wall term cancel.
All quantities use a common finite active box with frozen exterior
fields, specified in Sec.~\ref{app:dtO4-proof}. Subtract first, take the
volume to infinity at fixed $x$, and then take $x\to\infty$, separately
for every field coefficient.

Let $F_n[s]$ be the real-space diagonal matrix element of the order-$n$
canonical reduction in background $s$, with units $J^{1-n}$:
\begin{equation}
\mathcal E_{\rm diag}^{[4]}[s;h]
=F_0[s]+h^2F_2[s]+h^4F_4[s],
\label{dtO4:energy-expansion}
\end{equation}
with $F_0[s]=-J\sum_{i<j}s_is_j/|i-j|^4$ understood in the same
finite-volume convention. The retained projectors are the two-wall sector for $D_x$
and the corresponding one-wall or vacuum sectors for the other
backgrounds. The scalar coefficients include the cores and use the full
sector projectors, rather than rank-one perturbation theory for a
movable wall; they differ from the matrices $H_n$. Odd diagonal orders vanish
because an odd number of flips cannot return to the same spin
configuration, although odd-order hopping remains in the matrix.
The connected diagonal interaction is
\begin{equation}
\mathcal I_x\mathcal E_{\rm diag}^{[4]}[\cdot;h]
=\mathcal I_xF_0+h^2\mathcal I_xF_2+h^4\mathcal I_xF_4.
\label{dtO4:connected-expansion}
\end{equation}

\paragraph{Linked virtual processes.}
In a classical two-wall background $s_i=\pm1$, use the same gaps
$\Delta=4J\zeta(4)$ and $\Delta_2(r)=2\Delta-4J/r^4$ as in the
magnetization calculation. Remove the same fixed neighborhoods of radius
eight about $0$ and $x$ in all four backgrounds, and call the remaining
sites bulk sites. The corresponding local gaps are
\begin{equation}
d_i=2Js_i\sum_{j\ne i}\frac{s_j}{|i-j|^4},\qquad k_{ij}=\frac{4Js_is_j}{|i-j|^4},\qquad
d_{ij}=d_i+d_j-k_{ij}.
\label{dtO4:gaps}
\end{equation}
The first gap follows by reversing all bonds incident to $i$;
subtracting their shared bond gives $d_{ij}$, just as for the vacuum.
For bulk sites the retained active subspace is rank one. The resolvent
denominators are $-d_i$ and $-d_{ij}$, so the second-order term is
$-1/d_i$. The one-site fourth-order normalization term is $1/d_i^3$.
For two sites, the four walks have intermediate flip sets
$\{i\},\{i,j\},\{i\}$,
$\{i\},\{i,j\},\{j\}$, and their $i\leftrightarrow j$ partners.
Their sum is
$-(d_i^{-1}+d_j^{-1})^2/d_{ij}$; separated bulk flips in $D_x$
include a six-wall midpoint. Walks returning to the retained state
at the midpoint enter the folded normalization term rather than an
inverse zero denominator. Its mixed part is
$d_i^{-2}d_j^{-1}+d_i^{-1}d_j^{-2}$.
Thus for generic gaps $a,b$ and signed bond $k$ their linked sum is
\begin{equation}
\Phi(a,b,k)
=\frac{a+b}{a^2b^2}-\frac{(a+b)^2}{a^2b^2(a+b-k)}=-\frac{k(a+b)}{a^2b^2(a+b-k)}.
\label{dtO4:phi}
\end{equation}
The disconnected part therefore cancels identically for $k=0$, including
unequal gaps. The coefficients in Eq.~\eqref{dtO4:energy-expansion} have the form
\begin{equation}
\begin{split}
F_2[s]&=-\sum_{i\,{\rm bulk}}d_i^{-1}+F_{2,{\rm core}}[s],\\
F_4[s]&=\sum_{i\,{\rm bulk}}d_i^{-3}
+\sum_{\substack{i<j\\i,j\,{\rm bulk}}}\Phi(d_i,d_j,k_{ij})
+F_{4,{\rm core}}[s].
\end{split}
\label{dtO4:functional}
\end{equation}
In the uniform vacuum, $d_i=\Delta$ and
$d_{ij}=\Delta_2(|i-j|)$. Consequently the coefficient of $h^4$ in the
vacuum energy density is
\begin{equation}
\frac1{\Delta^3}
+\sum_{r\ge1}\Phi(\Delta,\Delta,4J/r^4)
=\frac{B}{\Delta^2},
\label{dtO4:vacuumB}
\end{equation}
recovering Eq.~\eqref{app:vacO4} from the same linked processes.
The residual $F_{n,{\rm core}}$ contains every process touching a
removed neighborhood, including pairs with one distant site, and uses
the canonical sector projector. Its contribution is controlled by the
lemma below.

\paragraph{Bulk response to the second wall.}
Write $\delta_i[s]=d_i[s]-\Delta$. Each opposite-spin bond contributes
$-4J/|i-j|^4$ to the gap change at each endpoint. Only bonds joining
$i\leq0$ to $j>x$ survive the four-background subtraction: their signs
are $+,-,-,+$ in $D_x,\mathsf L,\mathsf R,+$. At range $r>x$ there
are $r-x$ such bonds, giving
\begin{align}
T_x&=\sum_{r>x}\frac{r-x}{r^4}
=\frac1{6x^2}-\frac1{12x^4}+\cO(x^{-6}),\notag\\
\mathcal I_x\sum_i\delta_i&=16JT_x.
\label{dtO4:fieldsum}
\end{align}
The gap-sum identity is exact before core regularization. The expansion
follows from $T_x=\zeta(3,x+1)-x\zeta(4,x+1)$ and
Eq.~\eqref{app:Hurwitz}.

\paragraph{Auxiliary lemma (subleading terms).}
For the common radius-eight cores and $x>32$, the connected core
coefficients satisfy
$\mathcal I_xF_{n,{\rm core}}=\cO(J^{1-n}x^{-3})$, $n=2,4$.
The off-diagonal interaction has absolute hopping moments
\begin{equation}
\sum_{x'\ne x}|x'-x|^m| (\mathcal W_K^{[4]})_{xx'}|
=\cO(Jx^{-3}),\qquad m\ge1,
\label{dtO4:moments}
\end{equation}
for each fixed integer $m$ and fixed $u$ in the truncated polynomial,
with the bounds understood coefficientwise. In the bulk terms, the
linear gap response and the uniform-gap signed-bond contribution
determine the connected result up to $\cO(x^{-3})$ at each order.
Extending the bulk representatives through the common cores with gaps
set to $\Delta$ has the same subleading effect. These bounds hold in the stated order of volume and
separation limits. The proof is given in Sec.~\ref{app:dtO4-proof}.

For a scalar function $f$ of a local gap, the linear response is
$f'(\Delta)\delta_i$. We use $f(d)=-d^{-1}$ at second order and
$f(d)=d^{-3}$ for the repeated-site fourth-order term. The lemma
controls the summed nonlinear remainder, so Eq.~\eqref{dtO4:fieldsum}
gives
\begin{equation}
\mathcal I_x\sum_i f(d_i)
=16Jf'(\Delta)T_x+\cO(x^{-3}).
\label{dtO4:scalar}
\end{equation}
Dimensionful derivative bounds are included in the remainder constant.
The spatially summed linear response is therefore of order $x^{-2}$,
even though the effect of the distant wall at any fixed core site is
only of order $x^{-3}$.

For the pair term, write $k_r=4J/r^4$ and
$\Phi_a(k)=\partial_a\Phi(a,b,k)|_{a=b=\Delta}$. Its linear gap
response is $\Phi_a(k)(\delta_i+\delta_j)$. The lemma allows us to
use the positive vacuum bond $k_r$ in this response, up to subleading
terms. Each site is the endpoint of two pairs at every distance, so the
coefficient of $\sum_i\delta_i$ in $F_4$ is
\begin{equation}
-\frac3{\Delta^4}+2\sum_{r\geq1}\Phi_a(4J/r^4)
=\left.\partial_\Delta\frac{B}{\Delta^2}\right|_{\{k_r\}}.
\label{dtO4:gap-response}
\end{equation}

\paragraph{Direct bond-sign contribution.}
The uniform-gap pair term also changes when a bond changes sign. The
surviving exterior pairs give
\begin{align}
&\mathcal I_x\sum_{i<j}\Phi(\Delta,\Delta,k_{ij})\notag\\
&\quad=2\sum_{i\leq0,j>x}
 [\Phi(\Delta,\Delta,k_{j-i})-\Phi(\Delta,\Delta,-k_{j-i})]\notag\\
&\quad=-16J\Delta^{-4}T_x+\cO(J^{-3}x^{-6}).
\label{dtO4:direct}
\end{align}
Indeed $\partial_k\Phi(\Delta,\Delta,0)=-\Delta^{-4}$; the outer
factor two and the sign difference give $-4k_r\Delta^{-4}$ per pair.
A quadratic Taylor remainder suffices, since
$\sum_{r>x}(r-x)r^{-8}=\cO(x^{-6})$.

\paragraph{Tail coefficient in terms of $B$ and $B'$.}
The zeroth-order coefficient gives
\begin{equation}
\mathcal I_xF_0=V_4^{(0)}(x)-V_4^{(0)}(\infty)
=-4JT_x,
\label{dtO4:F0tail}
\end{equation}
which is the exact Hurwitz difference in Eq.~\eqref{app:Hurwitz}.
At second order,
Eq.~\eqref{dtO4:scalar} gives
\begin{equation}
\mathcal I_xF_2=\frac{16J}{\Delta^2}T_x
+\cO(J^{-1}x^{-3}).
\label{dtO4:F2tail}
\end{equation}
At fourth order, the coefficient of the summed linear gap change is
the derivative of the vacuum expression in Eq.~\eqref{dtO4:vacuumB}.
Here the bonds $k_r=4J/r^4$ are held fixed, so that
$\Delta_2(r)=2\Delta-k_r$. With the longitudinal-field convention used
in Eq.~\eqref{app:mzO4}, $\partial_\ell=2\partial_\Delta$ at fixed
$\{k_r\}$, and therefore
\begin{equation}
\left.\frac{\partial}{\partial\Delta}
\frac{B}{\Delta^2}\right|_{\{k_r\}}
=\frac{B'}{2\Delta^2}-\frac{2B}{\Delta^3}.
\label{dtO4:Bderivative}
\end{equation}
Combining this response with the direct pair contribution
$-\Delta^{-4}$ from Eq.~\eqref{dtO4:direct} yields
\begin{equation}
\mathcal I_xF_4
=16J\left(\frac{B'}{2\Delta^2}
-\frac{2B}{\Delta^3}-\frac1{\Delta^4}\right)T_x+\cO(J^{-3}x^{-3}).
\label{dtO4:F4tail}
\end{equation}
The last term in parentheses is required: changing the background changes
both the single-spin gaps and the sign of the bond linking two virtual
flips. Keeping only the gap response would miss this contribution.

Since $T_x=1/(6x^2)+\cO(x^{-4})$ and the lemma bounds the
off-diagonal interaction, the complete row sum is
\begin{align}
\sum_{x'}(\mathcal W_K^{[4]})_{xx'}
&=-\frac{A^{[4]}}{x^2}
+\sum_{n=0}^4 u^n\cO_n(Jx^{-3}),\notag\\
A^{[4]}&=\frac{2J}{3}\left[
1-\frac{4h^2}{\Delta^2}
+h^4\left(\frac4{\Delta^4}
+\frac{8B}{\Delta^3}-\frac{2B'}{\Delta^2}\right)\right]\notag\\
&=\frac{2J}{3}m_z^{2,[4]}.
\label{dtO4:matching}
\end{align}
The equality follows by comparison with Eq.~\eqref{app:mz2B}, after
evaluating the interaction rather than inserting the magnetization
into its definition. Numerically,
\begin{equation}
\frac{A^{[4]}}{J}
=\frac23-0.1422770287322u^2-0.0353497401259u^4.
\label{dtO4:coefficients}
\end{equation}
Equation~\eqref{dtO4:moments} also excludes another marginal derivative
term: every higher hopping moment is subleading when acting on a slowly
varying envelope. The coefficient $A^{[4]}$ is independent of each fixed
$K$ in the centered gauge. For an isolated channel with a nondegenerate
minimum at $q=0$ and $D_K>0$, it supplies the interaction in
Eq.~(4), together with
$D_K^{[4]}\equiv[D_K]_{\leq4}$.

The calculation is coefficientwise through fourth order; it neither
proves convergence of the full weak-field series nor gives uniform
control near a degenerating threshold or the bulk critical point.
Within this scope, the complete canonical kernel reproduces
Appendix~\ref{app:tail}'s dressed-tail coefficient, with the core and
separation-dependent hopping controlled through the retained order.
This connects the microscopic interaction to $V_{\rm rel}$ in the
threshold theory without identifying its core with $V_{\rm corr}$ or
requiring a nonzero thermodynamic bare two-wall overlap.

\subsection{Control of subleading terms in the dressed tail}
\label{app:dtO4-proof}

We prove the auxiliary lemma of Sec.~\ref{app:dtO4}.

\paragraph{Finite-volume prescription.}
Use a common active box $\Lambda_M=\{-M,\ldots,x+M\}$, with transverse
flips allowed only there and exterior spins frozen in each background.
Retain the exterior Ising fields and count once each classical bond with
an endpoint in the box, in all four backgrounds.
Form Eq.~\eqref{dtO4:connected-subtraction} at finite $M$, then take
$M\to\infty$ at fixed $x$ before taking $x\to\infty$, coefficientwise.
Convergence of the full weak-field series is not assumed.

\paragraph{Canonical cancellation and core terms.}
Assign each spin an independent transverse field $h_i$. A monomial of
degree at most four involves at most four active sites; freeze the rest
while retaining their exact Ising fields. For $x>32$, the cores are
disjoint and unchanged segments on either side prevent annihilation of
the interval. Exactly two retained walls then require one per core and
no extra pair at a bulk spectator. The active retained space is therefore
the product of local wall-position spaces and a rank-one spectator
space; the full two-wall projector need not factorize.

If $\Pi$ is the continued invariant subspace, the canonical isometry is
\begin{equation}
\begin{split}
\mathcal V_{\rm can}&=\Pi P(P\Pi P)^{-1/2},\\
[\mathcal V_{\rm can}]_{\leq4}&=[\ee^{-S}P]_{\leq4}.
\end{split}
\label{dtO4:isometry}
\end{equation}
For each finite active problem near zero field, $\|\Pi-P\|<1$ ensures
the positive inverse square root exists on the retained range. When the
active parts decouple, both $P$ and $\Pi$ are products; hence $P\Pi P$,
its inverse square root, and the isometry factorize. The effective
Hamiltonian is additive, so all mixed field coefficients cancel,
including folded and normalization terms. This uses multiplicativity
on the retained range~\cite{Bravyi2011SW}, not of the full unitary.

The denominator bound can be made independent of separation. For these
far-separated templates an eliminated state has at least four walls,
whereas a retained state has two, giving a nearest-neighbor gap of at
least $4J$ (the one-wall and vacuum sectors have the same two-wall
increment). For a change on a set of $m\leq4$ spins, bounding each
affected non-nearest-neighbor bond in absolute value gives at most
$4mJ\sum_{r\geq2}r^{-4}$. Thus each required denominator is at least
\begin{equation}
g_* = 4J-16J[\zeta(4)-1]>2.68J.
\label{dtO4:gap-bound}
\end{equation}
The same estimate holds while turning off long-range bonds between
active parts. At a fixed order there are finitely many walks on at most
four active sites. Their rational coefficients and the required gap
derivatives are uniformly bounded by powers of $g_*^{-1}$.
If $c(\mathbf k)$ is a mixed coefficient and $\mathbf k$ the vector
of bonds joining two active parts, additivity gives $c(0)=0$, and
\begin{equation}
c(\mathbf k)=\int_0^1\mathbf k\cdot
\nabla c(t\mathbf k)\,dt.
\label{dtO4:cut-bound}
\end{equation}
It is therefore bounded by the sum of these connecting bonds, with the
frozen-spin fields held fixed. The argument also bounds derivatives
with respect to those fields.

In a diagonal fourth-order term the active sites are either one site
flipped four times or two sites each flipped twice. In an off-diagonal
term, every oddly flipped site is within four sites of a boundary,
so there is at most one unbounded, twice-flipped spectator. By
Eq.~\eqref{dtO4:cut-bound}, its mixed coefficient is bounded by
$C(1+|n|)^{-4}$ at distance $n$ from a core, with the appropriate power
of $J$ included in $C$. For $|n|\leq x/2$, the distant interface
changes its frozen-spin field by $\cO(Jx^{-3})$.
For $|n|>x/2$, sum the absolute spectator weights directly. The two
ranges are bounded by
\begin{equation}
x^{-3}\sum_{|n|\leq x/2}(1+|n|)^{-4}
+\sum_{|n|>x/2}(1+|n|)^{-4}=\cO(x^{-3}).
\label{dtO4:spectator-bound}
\end{equation}
This also controls diagonal core--bulk pairs, including spectators near
the other wall, where the field change need not be small. Terms coupling
the two fixed cores have connecting bonds $\cO(Jx^{-4})$; attaching a
spectator to its nearer core gives the same summable bound. Terms confined
to one core change by $\cO(Jx^{-3})$. Thus the connected core terms and
the absolutely summed off-diagonal interaction are $\cO(x^{-3})$,
with the respective powers of $J$ understood. Core-gap regularization
has the same bound.

Away from the cores, a path with at most four flips cannot connect
separations with $|x'-x|>4$. The center displacement is bounded as well,
and the centered phases in Eq.~\eqref{app:HnKO4} have modulus one.
The absolute off-diagonal bounds therefore give every fixed moment in
Eq.~\eqref{dtO4:moments}.

\paragraph{Spatial Taylor remainders.}
The local connected gap change in Eq.~\eqref{dtO4:fieldsum} is
$8J\sum_{j>x}(j-i)^{-4}$ at $i\leq0$, zero for $1\leq i\leq x$,
and the reflected value for $i>x$. Regularizing the finitely many core
sites therefore changes its sum only by $\cO(Jx^{-3})$.

For a single interface, $|\delta_i|\leq CJ\tau(i)$, where
$\tau(n)=(1+|n|)^{-3}$ with a fixed unit shift if needed at the wall.
This follows by bounding the half-line sum of $r^{-4}$.
Set $w(n)=(1+|n|)^{-4}$. Splitting each convolution at $x/2$,
one factor is $\cO(x^{-3})$ while the other is summable, giving
\begin{align}
\sum_n\tau(n)\tau(n-x)&=\cO(x^{-3}),\notag\\
(w*\tau*\tau)(x)&=\cO(x^{-3}),\notag\\
\sum_{i\leq0,j>x}\frac{\tau(i)+\tau(j-x)}{(j-i)^4}
&=\cO(x^{-3}).
\label{dtO4:bounds}
\end{align}
For the last bound, first sum over the far endpoint, obtaining
$\cO[(x+|i|)^{-3}]$, and then use $\sum_i\tau(i)<\infty$.

The functions $f(d)=-d^{-1}$ and $d^{-3}$ used in
Eq.~\eqref{dtO4:scalar} require bounded first and second derivatives
on the entire interval sampled by the gaps and their Taylor
interpolations. Outside the common radius-eight cores, for $x>32$, the gap changes
are bounded by the corresponding tails of $\sum r^{-4}$. All these
arguments lie in $[3J,5J]$; the prescribed core value $\Delta$ lies
there too. Thus the required derivatives of these two functions are
bounded. Likewise, for $a,b$ in this interval and
$|k|\leq4J/r^4$, $a+b-k\geq2J$, so all required derivatives of
$\Phi$ are bounded and its gap derivatives are $\cO(|k|)$.

To see the scalar response explicitly, set
$a_i=\delta_i[\mathsf L]$, $b_i=\delta_i[\mathsf R]$, and
$c_i=\mathcal I_x\delta_i$. Before core regularization,
$\delta_i[D_x]=a_i+b_i+c_i$,
$\|c\|_\infty=\cO(Jx^{-3})$, and $\sum_i|c_i|=16JT_x$.
A two-variable Taylor remainder bounds
\begin{align}
&\mathcal I_x f(d_i)-f'(\Delta)c_i\notag\\
&\quad=\cO\!\left(|a_i b_i|
+|c_i|(|a_i|+|b_i|+|c_i|)\right),
\label{dtO4:scalar-remainder}
\end{align}
with dimensionful derivative bounds included in the constant.
Summing with Eq.~\eqref{dtO4:bounds} gives
Eq.~\eqref{dtO4:scalar}. Common core regularization preserves this
result: nonlinear field-tail overlaps are subleading, but the summed
linear response is not.

For the pair term, subtract the first-order gap expansion
$\Phi(\Delta,\Delta,k)+\Phi_a(k)(\delta_i+\delta_j)$.
After connected subtraction, the quadratic remainders contain
products of the two interface tails weighted by $w(i-j)$, bounded by
Eq.~\eqref{dtO4:bounds}. Replacing the signed-bond response
$\Phi_a(k_{ij})$ by $\Phi_a(4J/|i-j|^4)$ produces only crossing-bond
terms: $|\Phi_a(k)-\Phi_a(-k)|\leq C|k|$. Their connected part is
bounded by the same tail overlaps or the exterior-pair sum in
Eq.~\eqref{dtO4:bounds}. These errors are $\cO(x^{-3})$.
Together these estimates establish the bulk reduction used in
Eq.~\eqref{dtO4:gap-response}. The absolute summable bounds justify
enlarging the active box at fixed $x$ and then taking the separation
limit coefficientwise. This completes the proof of the lemma.

\paragraph{Accuracy of truncated threshold estimates.}
Using the truncated amplitude and curvature, we estimate
$s_K\simeq\sqrt{A^{[N]}/D_K^{[N]}-1/4}$ for $N=2,4$.
We evaluate this ratio and square root directly. This retains the
competition between hopping harmonics when the leading curvature is
small, but it does not determine the higher field powers generated by
these operations. For omitted corrections $\delta A,\delta D_K$,
the leading propagated errors are
\begin{equation}
\delta g_K\simeq\frac{\delta A}{D_K}
-\frac{A\,\delta D_K}{D_K^2},\qquad
\delta s_K\simeq\frac{\delta g_K}{2s_K}.
\label{dtO4:ratio-errors}
\end{equation}
They are small only if the omitted curvature is small relative to
$D_K$ and $|\delta g_K|\ll s_K^2$. No uniform error order follows
near $D_K=0$ or $g_K=1/4$ merely from using fourth-order inputs.

\subsection{Consistent size operators and system-size dependence}

To compare with Eq.~\eqref{app:EDsize}, transform each physical observable with the same generator:
\begin{equation}
O_{\rm eff}^{[N]}=\left[P\ee^SO\ee^{-S}P\right]_{\leq N},
\qquad O=P_2,X,X^2.
\label{app:OeffO4}
\end{equation}
The effective eigenvectors have the identity metric because the canonical transformation is unitary. With $T=QS_1P$ and $G=T^\dagger T$, the second-order terms are
\begin{align}
(P_2)_{\rm eff}^{[2]}&=I-h^2G,\notag\\
(X^p)_{\rm eff}^{[2]}&=X^p-\frac{h^2}{2}\{G,X^p\},\quad p=1,2.
\label{app:sizeoperatorsO2}
\end{align}
Through either retained order, $W_{2,n}=\langle(P_2)_{\rm eff}\rangle_n$ and $\xi_n^2=\langle(X^2)_{\rm eff}\rangle_n/W_{2,n}$.

The bare two-wall weight includes vacuum fluctuations throughout the
ring. It can therefore decrease with $L$ even when a dressed meson
remains well defined; its finite-order expansion is not uniform at fixed
$h$ as $L\to\infty$. Conditional sizes require consistent numerator and
denominator dressing and an order/size check. The $L=256$ curves in
Fig.~3 are an effective-model extension, whose accuracy
includes the truncation of both the Hamiltonian and the observable.

For a vacuum-subtracted separation kernel, the row-sum diagnostic
$x^2[E_c-\sum_{x'}H_{xx'}]$ probes the tail after bulk motion has been
separated and the regime $1\ll x\ll L$ has been reached.

\subsection{Independent validation against QuSpin}
\label{app:quspin-validation}

The second- and fourth-order effective-Hamiltonian implementations are
benchmarked against a separate full-spin ED calculation using QuSpin~\cite{Weinberg2017QuSpin}. The two calculations use independent code
paths: QuSpin constructs the full spin Hamiltonian, symmetry basis, and
physical observables directly, whereas the perturbative calculation uses
the Schrieffer--Wolff recursion above. Both use Pauli operators,
Eq.~(2), and the same periodized coupling in
Eq.~\eqref{app:periodic_coupling}.

The benchmark uses $L=16$, $\alpha=4$, $k=4$ ($K=\pi/2$), and
spin-flip parity $z=+1$. The QuSpin target sector has dimension 2065, the
$K=0$, $z=+1$ ground-state sector has dimension 2068, and the two-wall
effective sector has dimension eight. We computed 12 ED eigenstates in
the target sector at 25 logarithmically spaced fields
$0.006\leq u=h/J\leq0.04$. Hermiticity and symmetry checks pass, and the
largest ED residual is $2.22\times10^{-13}J$.

Excitation energies $e_n=E_n-E_{\rm vac}$ subtract the ED ground-state energy or the corresponding-order perturbative vacuum energy. One-to-one matching maximizes normalized overlaps with the two-wall projections of ED states. For the lowest meson branch, $n=0$, we compare $|e_0^{[N]}-e_0^{\rm ED}|/J$, $|W_{2,0}^{[N]}-W_{2,0}^{\rm ED}|$, and $|\xi_0^{[N]}-\xi_0^{\rm ED}|/\xi_0^{\rm ED}$. The physical weight and conditional size are defined in Eq.~\eqref{app:EDsize}; their perturbative values use the consistently transformed operators in Eq.~\eqref{app:OeffO4}.

We also compare the complete matrices in the canonical gauge~\cite{Bravyi2011SW}. Let $C$ embed the bare two-wall basis into the QuSpin symmetry basis and let $\Psi$ contain the eight ED eigenstates adiabatically connected to that band. From the singular-value decomposition
\begin{equation}
C^\dagger\Psi=U\Sigma V^\dagger,\qquad R=UV^\dagger,
\label{app:validation_polar}
\end{equation}
we obtain
\begin{align}
H_{\rm can}^{\rm ED}&=R\,\mathrm{diag}(E_n^{\rm ED})R^\dagger,\notag\\
O_{\rm can}^{\rm ED}&=R(\Psi^\dagger O\Psi)R^\dagger,
\label{app:validation_canonical}
\end{align}
for $O=P_2,X,X^2$, with $X$ extended by zero outside the two-wall subspace. This construction requires $C^\dagger\Psi$ to have full rank and the retained band to remain separated from eliminated states. Spectral-norm differences from the order-$N$ polynomials test all matrix elements, with generic remainders $\cO(u^{N+1})$ (in units of $J$ for the Hamiltonian).

For even $L$, $\mathcal U=\prod_i\sigma_i^z$ maps $h$ to $-h$ while preserving $K$, $z$, and the measured observables. A branch nondegenerate at $h=0$ and continued analytically therefore has even energy and observable expectation values. Under these conditions, scalar errors start generically at $u^4$ after $\mathcal{O}2$ and $u^6$ after $\mathcal{O}4$, whereas the full matrix errors start at $u^3$ and $u^5$.

\begin{table}[!tbp]
\centering
\begin{tabular}{lcc}
\toprule
Error & $\mathcal{O}2$ & $\mathcal{O}4$\\
\midrule
Excitation energy & 4.1236 & 6.0424\\
$W_2$ & 4.0483 & 5.9844\\
Relative $\xi$ & 4.7565 & 5.8753\\
\midrule
$H_{\rm eff}$ matrix & 3.0006 & 5.0026\\
$(P_2)_{\rm eff}$ matrix & 3.0090 & 5.0049\\
$X_{\rm eff}$ matrix & 3.0085 & 5.0048\\
$(X^2)_{\rm eff}$ matrix & 3.0079 & 5.0038\\
\bottomrule
\end{tabular}
\caption{QuSpin validation at $L=16$, $K=\pi/2$, $z=+1$, and $\alpha=4$: fitted powers $p$ in errors proportional to $u^p$. Scalar errors refer to the lowest meson branch. Their fits use nine resolved samples in $0.01\leq u\leq0.02$ (sample endpoints $0.0104342$ and $0.0196379$), except for the $\mathcal{O}4$ size fit, which uses seven. Matrix errors are spectral norms over the full eight-state effective sector.}
\label{tab:validation}
\end{table}

\begin{figure}[!tbp]
\centering
\includegraphics[]{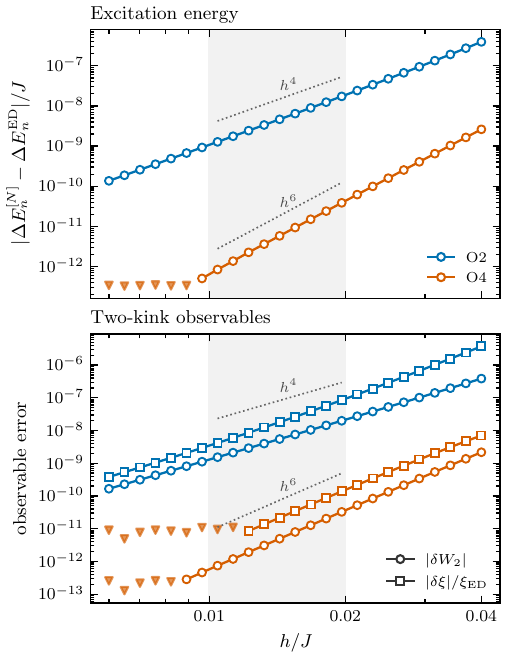}
\caption{Independent QuSpin validation at $L=16$, $K=\pi/2$, $z=+1$, $n=0$~(lowest meson), and $\alpha=4$, with periodized couplings. Top: absolute excitation-energy error for the lowest meson branch in units of $J$; $\Delta E_n$ in the panel denotes excitation energy above the vacuum. Bottom: absolute error in the physical two-wall weight $W_2$ (circles) and relative error in the conditional rms separation $\xi$ (squares). Blue and orange denote $\mathcal{O}2$ and $\mathcal{O}4$. Dotted lines show $h^4$ and $h^6$ powers with arbitrary normalizations. Shading marks the fit window $0.01\leq h/J\leq0.02$; downward triangles are resolution-limited points excluded from the fits.}
\label{fig:quspin-validation}
\end{figure}

Figure~\ref{fig:quspin-validation} and Table~\ref{tab:validation} support
the expected perturbative orders, with particularly clear $u^3/u^5$
matrix remainders. Energy and weight fits are close to $u^4/u^6$; size
fits show larger finite-window deviations. The $\mathcal{O}2$ size
exponent $4.7565$ illustrates this finite-window drift. Errors at
residual-based or floating-point resolution limits are excluded from
fits, leaving seven resolved points for the $\mathcal{O}4$ size. This
benchmark checks the finite-ring Hamiltonian and dressed observables;
the large-$L$ scaling and long-distance tail require the separate limits
described above.

\section{Analytic one-kink dispersion through fourth order}
\label{app:fourth_order}

The isolated-wall version of the same perturbation theory gives
\begin{equation}
\varepsilon(p)=\varepsilon_0-2\sum_{r\geq1}t_r\cos(rp).
\label{app:kinkdisp}
\end{equation}
The transformation that changes $h$ to $-h$ shifts the one-kink momentum by $\pi$, so $t_r$ has the parity of $r$ in $h$. At $\alpha=4$,
\begin{align}
t_1/J&=u-bu^3+\cO(u^5),& t_2/J&=u^2/4+du^4+\cO(u^6),\notag\\
t_3/J&=cu^3+\cO(u^5),& t_4/J&=eu^4+\cO(u^6).
\label{app:dispersionO4}
\end{align}
The offset $\varepsilon_0$ is fixed by the vacuum-subtracted diagonal kernel. For example, with $\Delta_{\rm w}(n)=4J\sum_{r=1}^n r^{-\alpha}$,
\begin{equation}
\begin{split}
\varepsilon_0={}&2J\zeta(\alpha-1)+\frac{2h^2}{\Delta}\\
&+2h^2\sum_{n=1}^{\infty}
\left(\frac1\Delta-\frac1{\Delta_{\rm w}(n)}\right)+\cO(h^4).
\end{split}
\label{app:kinkselfenergy}
\end{equation}
The bracket is $0.407782748943/J$ at $\alpha=4$. Its fourth-order counterpart follows from the same SW recursion and is unnecessary for curvature or spacing ratios.

\subsection{Explicit hopping coefficients}

Set $\mathsf H_n=\sum_{r=1}^n r^{-4}$. The nearest-neighbor renormalization is the convergent sum
\begin{equation}
b=\frac1{16}\left[1+\frac1{\zeta(4)^2}
+2\sum_{n=1}^{\infty}\frac{\mathsf H_{n+1}-\mathsf H_n}{\mathsf H_n^2\mathsf H_{n+1}}\right]=0.125224434014537\ldots .
\label{app:b}
\end{equation}
The third-neighbor hop involves only local denominators. With $\Delta_{\rm w}(1)=4J$ and $\Delta_{\rm w}(2)=4J\mathsf H_2$,
\begin{align}
t_3&=h^3\left[\frac2{\Delta_{\rm w}(1)\Delta_{\rm w}(2)}+\frac1{\Delta_{\rm w}(2)^2}
-\frac1{\Delta_{\rm w}(1)^2}\right],\notag\\
c&=\frac{511}{4624}=0.110510380622837\ldots .
\label{app:c}
\end{align}

For the fourth-order correction to $t_2$, retain the local core and both nonlocal excursion families. Define
\begin{equation}
\mathcal F(x,y,z,q)=\frac1{qz}+\frac1q+\frac1{xyz}
+\frac1{xqz}+\frac1{xq}-\frac1{x^2y}-\frac1{xy^2}+\frac1{x^3}-\frac1x-\frac1{x^2},
\label{eq:app_F}
\end{equation}
With $q_n^-=1+\mathsf H_n-(n-1)^{-4}$ and $q_n^+=1+\mathsf H_n+(n+2)^{-4}$, the two families are
\begin{align}
W_n^{(-)}&=\mathcal F(\mathsf H_n,\mathsf H_{n-1},\mathsf H_{n-2},q_n^-),\quad n\geq3,\notag\\
W_n^{(+)}&=\mathcal F(\mathsf H_n,\mathsf H_{n+1},\mathsf H_{n+2},q_n^+),\quad n\geq1.
\label{app:excursions}
\end{align}
In units of $J$, the exceptional core denominators are $a_1=4$ and $a_2=17/4$, so
\begin{equation}
d_{\rm core}=\frac1{a_1^3}-\frac2{a_1^2a_2}-\frac2{a_1a_2^2}
-\frac1{a_2^3}=-\frac{17135}{314432}.
\label{eq:app_dcore_general}
\end{equation}
The complete coefficient is
\begin{equation}
d=-\frac{17135}{314432}
+\frac1{64}\left[\sum_{n=3}^{\infty}W_n^{(-)}
+\sum_{n=1}^{\infty}W_n^{(+)}\right]=-0.051976389022667\ldots .
\label{eq:app_d}
\end{equation}
The separate lower limits prevent double counting of the exceptional configurations. The cancellation within each summand must be retained when evaluating the convergent tails.

The fourth-neighbor hop is local. Define dimensionless gaps $a=4$, $b_2=4\mathsf H_2$, $c_3=4\mathsf H_3$, and
\begin{equation}
e_3=4(1+2/16+1/81),\qquad g_3=4(2+1/81).
\end{equation}
Summing the four-flip paths and folded terms gives
\begin{align}
e={}&-\frac2{b_2^3}+\frac1{a^2g_3}+\frac1{b_2^2e_3}
+\frac1{c_3^2e_3}+\frac1{c_3^2g_3}-\frac1{ab_2^2}-\frac2{a^2b_2}+\frac2{b_2^2c_3}
+\frac2{ab_2c_3}+\frac2{ac_3g_3}+\frac2{b_2c_3e_3}\notag\\
={}&\frac{4461217431258369}{73296639439215808}
=0.060865238370963\ldots .
\label{app:e}
\end{align}
As a check on both the local and nonlocal terms, the nearest-neighbor
limit gives $b=1/8$, $c=1/8$, $d=-1/16$, and $e=5/64$.
These values follow independently by Fourier expanding the exact
nearest-neighbor dispersion
$\varepsilon(p)=2J\sqrt{1+u^2-2u\cos p}$ through $u^4$.
The vacuum formulas also reduce to
$m_z^2=(1-u^2)^{1/4}=1-u^2/4-3u^4/32+\cO(u^6)$.

\subsection{Threshold coefficients and the quartic-zero contour}

With $p=K/2$, differentiation of Eq.~\eqref{app:kinkdisp} gives
\begin{align}
E_c(K)&=2\varepsilon_0-4\sum_r t_r\cos(rp),\notag\\
D_K&=2\sum_r r^2t_r\cos(rp),\notag\\
C_K&=-\frac16\sum_r r^4t_r\cos(rp),\notag\\
F_K&=\frac1{180}\sum_r r^6t_r\cos(rp).
\label{app:gradientsO4}
\end{align}
At second order, $D_K=2h\cos p+2h^2\cos(2p)/J$ and $C_K=-h\cos p/6-2h^2\cos(2p)/(3J)$. The virtual two-site hop has excitation energy $4J$; contributions from longer-range couplings cancel across an isolated wall, fixing $t_2=h^2/(4J)$.

The leading projection alone has $C_K=-D_K/12$ and cannot give $C_K=0$ with $D_K>0$. Dressing supplies the higher harmonics needed for the nontrivial root
\begin{equation}
t_1\cos p+16t_2\cos(2p)+81t_3\cos(3p)+256t_4\cos(4p)=0.
\label{app:CzeroO4}
\end{equation}
Let $\epsilon=\pi-K$ and $D_K^{\rm weak}=2h_0\cos(K/2)$. Expanding the root connected to $h=0$, $K=\pi$, yields
\begin{align}
\frac{h_0}{J}&=\frac\epsilon8+\gamma\epsilon^3+\cO(\epsilon^5),\notag\\
\gamma&=\frac18\left[\frac{11}{24}-\frac{b+243c}{64}
-\frac d{16}+e\right]=0.01261204535\ldots ,
\label{app:hzeroO4}\\
\frac{D_K}{D_K^{\rm weak}}&=\frac34+\beta\epsilon^2+\cO(\epsilon^4),\notag\\
\beta&=\frac{3(-b+45c-16e)}{256}=0.04539725250\ldots .
\label{app:DratioO4}
\end{align}
In particular,
\begin{equation}
D_K\sim\frac{3J\epsilon^2}{32},\qquad
s_K\sim\frac8{3\epsilon},\qquad (K\to\pi\text{ along }h_0).
\label{app:zone_edge_asymptotes}
\end{equation}
Although $t_2/t_1=\cO(u)$, on this contour $\cos(K/2)=\cO(u)$, so the
first two harmonics contribute to $D_K$ at the same order. Their
competition explains the finite $3/4$ curvature ratio in a weak-field
limit. The leading limits $h_0/J\sim\epsilon/8$ and
$D_K/D_K^{\rm weak}\to3/4$ hold for $\alpha>2$; the coefficients
$\gamma,\beta$ quoted above are for $\alpha=4$. The scaling of $s_K$
in Eq.~\eqref{app:zone_edge_asymptotes} additionally uses the marginal
inverse-square attraction.

\section{Choice of fields and ED--perturbation comparison}
\label{app:curvature_tuned}

\begin{figure}[!tbp]
\centering
\includegraphics[]{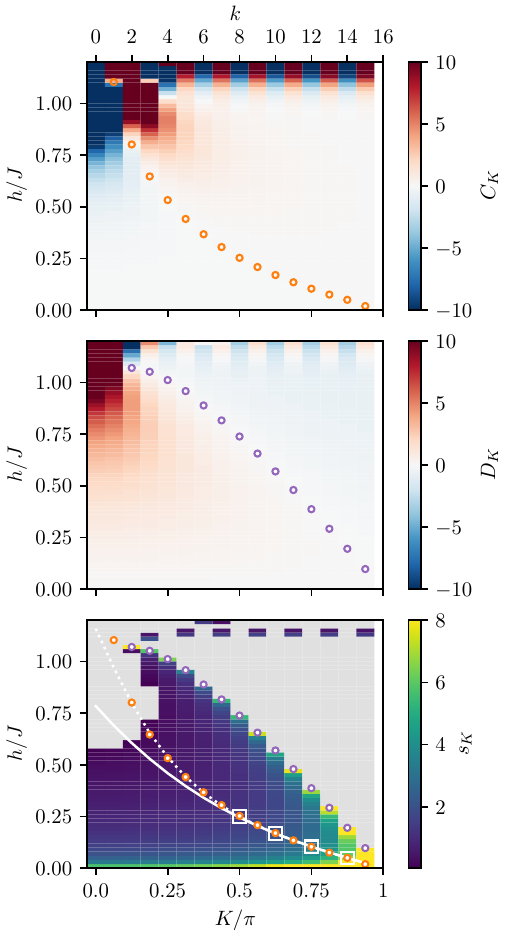}
\caption{Twisted-sector threshold scan for $L=32$, $\alpha=4$. From top to bottom: $C_K$, $D_K$, and $s_K$ versus $h/J$ and $K/\pi$, with $k=LK/(2\pi)$ on the top axis. Orange (violet) circles follow the estimated nontrivial $C_K=0$ ($D_K=0$) contour; white squares mark the four selected pairs in Table~\ref{tab:h0_curve}. In the lower panel the dotted and solid white curves are the empirical interpolation and fourth-order contour, respectively. The anomaly formula applies only to an isolated quadratic minimum with $D_K>0$ and $A/D_K>1/4$.}
\label{fig:CK_zero_scan}
\end{figure}

The threshold scan in Fig.~\ref{fig:CK_zero_scan} identifies fields where the quartic kinetic correction is small. The relevant contour is the nonzero root $C_K(h_0)=0$ with $D_K(h_0)>0$, continued from the zone-edge weak-field branch. The trivial root at $h=0$ is excluded. Close to this contour the threshold is
\begin{equation}
E_2(K,q)=E_c(K)+D_Kq^2+F_Kq^6+\cO(q^8),
\label{app:tuned_threshold}
\end{equation}
up to the residual fitted $C_Kq^4$ term. This improves access to quadratic-threshold scaling without tuning the inverse-square attraction or enforcing $g_K=1/4$.

For selecting the four ED parameter pairs, the scan was interpolated with $y=\sin(K/2)$ and
\begin{equation}
\frac{h_0^{\rm fit}(K)}J=
\frac{u_c\cos(K/2)}{1+a_f y+b_f y^2+c_f y^3},
\label{app:contour_fit}
\end{equation}
where $u_c\simeq1.1577$, $a_f\simeq1.6568$, $b_f\simeq2.7434$, and
$c_f=4u_c-1-a_f-b_f$. The denominator at $K=\pi$ is therefore $4u_c$,
enforcing the derived asymptote $h_0/J\sim(\pi-K)/8$.
The parameter $u_c$ fixes the empirical $K=0$ endpoint of the
interpolation. The validity of an isolated kink channel near that
endpoint must be assessed separately from the fit. The solid
fourth-order curve uses Eq.~\eqref{app:hzeroO4}.

\begin{table}[!tbp]
\centering
\setlength{\tabcolsep}{4pt}
\begin{tabular}{lcccc}
\toprule
$K/\pi$ & $1/2$ & $5/8$ & $3/4$ & $7/8$\\
\midrule
$h/J$ & $0.250254$ & $0.167850$ & $0.103863$ & $0.0497694$\\
$D_K/J$ & $0.3208$ & $0.1543$ & $0.0621$ & $0.0147$\\
$|C_K|/J$ & $8\!\times\!10^{-4}$ & $3\!\times\!10^{-4}$ & $5\!\times\!10^{-5}$ & $3\!\times\!10^{-6}$\\
$|F_K|/J$ & $10^{-2}$ & $3\!\times\!10^{-3}$ & $8\!\times\!10^{-4}$ & $2\!\times\!10^{-4}$\\
{$m_z^2$} & {$0.9864$} & {$0.9939$} & {$0.9977$} & {$0.9995$}\\
{$s_K$} & {$1.3417$} & {$2.0114$} & {$3.2334$} & {$6.7114$}\\
{$\Lambda_K$} & {$10.3974$} & {$4.7680$} & {$2.6422$} & {$1.5970$}\\
\bottomrule
\end{tabular}
\caption{Fields selected from the fitted contour and finite-stencil
threshold coefficients from $L=32$ ED. The residual $C_K$ measures the
deviation from the quartic-zero contour. Field values retain the
precision used for the figures; the printed fit parameters are rounded.
}
\label{tab:h0_curve}
\end{table}

Figure~3 uses all four pairs for the ED towers and the three
larger momenta for the effective-exponent comparison. At $L=32$ the
perturbative and full-spin calculations use identical couplings and the
same symmetry sector. The $\mathcal{O}4$ curves reproduce the ED trend and
its finite-size turnover. Increasing only the effective-model size to
$L=256$ adds threshold levels and exposes a wider interval in which the
energy and size estimators approach the same $s_K$, with the broadest
plateau at the largest selected momentum. The remaining convergence is
controlled by sextic dispersion, the interface core, interaction-tail
corrections, and observable truncation.

\begin{figure}[!tbp]
\centering
\includegraphics[]{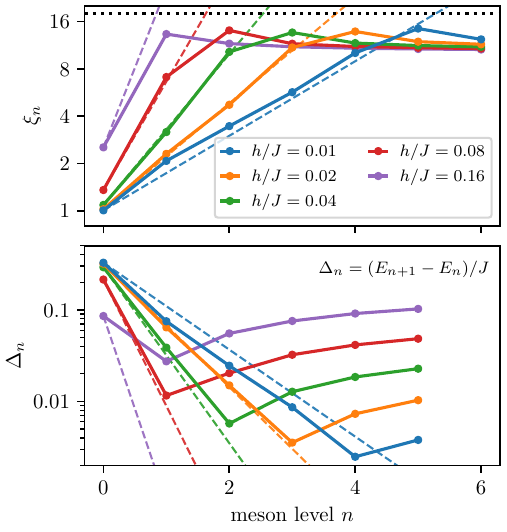}
\caption{Complementary $K=0$ ED benchmark at $L=36$ and $h/J=0.01,0.02,0.04,0.08,0.16$. Conditional rms sizes (top) and consecutive spacings $\Delta_n=(E_{n+1}-E_n)/J$ (bottom) are compared with the independently predicted geometric guide slopes. The dotted size line marks $L/2$. Saturation of the shallow-state sizes delimits the intermediate scaling window at these untuned parameters.}
\label{fig:K0_meson_scaling}
\end{figure}

The untuned $K=0$ data in Fig.~\ref{fig:K0_meson_scaling} give a
complementary finite-size comparison. Increasing the field steepens the
predicted size growth and makes finite-size bending appear at lower
level index. The useful states lie between the core-dominated levels
and this bending region. Testing both observables against independently
determined $m_z^2$ and $D_K$ identifies that window.

\section{Extracting the dressed tail without a threshold estimate}
\label{app:tail-difference}

We extract the inverse-square interaction from differences of two
Hamiltonian row sums. This eliminates the sensitivity to an uncertain
continuum threshold while retaining the complete dressed interaction.
Throughout this appendix, $\alpha=4$, and $[4]$ has the polynomial
meaning specified in Appendix~\ref{app:weakfieldO4}.

At fixed momentum $K$, define the vacuum-subtracted finite-ring row sum
\begin{equation}
R_{L,K}^{[4]}(x)=
\sum_{x'=1}^{L-1}
\left[H_{{\rm 2k},K}^{[4]}-E_{{\rm vac},L}^{[4]}I\right]_{xx'}.
\label{app:tail-rows}
\end{equation}
Here $H_{{\rm 2k},K}^{[4]}$ is the canonical two-kink Hamiltonian on a
ring of length $L$, and $E_{{\rm vac},L}^{[4]}$ is its corresponding
perturbative vacuum energy. We use the centered basis of
Eq.~\eqref{app:KxbasisO4}, with oriented separations $x=1,\ldots,L-1$,
before projection onto spin-flip parity. The sum includes every
diagonal and off-diagonal matrix element. A constant relative-coordinate
wavefunction represents relative momentum $q=0$ in this basis, so the
row sum includes both the diagonal interaction and the contribution
from separation-dependent hopping. Matrix row $x-1$ in zero-based
indexing carries the oriented separation $x$, not the folded observable
$X=\min(x,L-x)$. Rephasing or parity folding the basis changes a raw
row sum: after a basis transformation $U$, the same test vector would
be $U^\dagger\mathbf1$, rather than an untransformed vector of ones.

To connect the interaction calculation to this complete row sum, let
$\mathbf1$ denote the constant envelope in the centered basis.
Equation~\eqref{dtO4:interaction} implies
\begin{equation}
R_{\infty,K}^{[4]}(x)
=(H_{\rm free,K}^{[4]}\mathbf1)_x
+(\mathcal W_K^{[4]}\mathbf1)_x.
\label{app:tail-free-bridge}
\end{equation}
Away from the boundary, the first term is the free dispersion at
$q=0$, namely $E_c^{[4]}(K)$. Substituting the interaction row sum
from Eq.~\eqref{dtO4:matching} then gives
\begin{equation}
R_{\infty,K}^{[4]}(x)
=E_c^{[4]}(K)-\frac{A^{[4]}}{x^2}
+\mathcal O(Jx^{-3}),
\label{app:tail-row-asymptotic}
\end{equation}
where $E_c^{[4]}(K)$ is the two-independent-wall threshold at $q=0$
and $A^{[4]}$ is the dressed tail amplitude. The large-$x$ remainder
is understood coefficient by coefficient in the retained field
expansion. If the threshold estimate is
$\widetilde E_c^{[4]}=E_c^{[4]}+\delta E_c$, the direct diagnostic becomes
\begin{equation}
x^2\!\left[\widetilde E_c^{[4]}-R_{\infty,K}^{[4]}(x)\right]
=A^{[4]}+x^2\delta E_c+\mathcal O(J/x).
\label{app:tail-threshold-error}
\end{equation}
Thus even a small separation-independent threshold error grows
quadratically with $x$ and can obscure the asymptotic tail.

To remove this error, we compare two separations $x<y$ in the same
ring and at the same $K$ and $h$:
\begin{equation}
\widehat A_L(x,y)=
\frac{R_{L,K}^{[4]}(y)-R_{L,K}^{[4]}(x)}
{x^{-2}-y^{-2}}.
\label{app:tail-estimator}
\end{equation}
For a pure inverse-square tail this returns $A^{[4]}$ exactly,
independently of the threshold. Any common additive energy shift,
including a separation-independent vacuum-energy offset, also cancels.
The estimator is constructed directly from the kernel; no value of
$m_z^2$ or $A$ is inserted into it.

The remaining bias can be stated explicitly. Write the finite-ring
row sum as
$R_{L,K}^{[4]}(x)=C_{L,K}^{[4]}-A^{[4]}/x^2 + \varepsilon_{L,K}^{[4]}(x)$,
where $C_{L,K}^{[4]}$ is independent of separation and
$\varepsilon_{L,K}^{[4]}$ contains subleading interaction and finite-ring
corrections. Then
\begin{equation}
\widehat A_L(x,y)-A^{[4]}
=\frac{\varepsilon_{L,K}^{[4]}(y)-\varepsilon_{L,K}^{[4]}(x)}
{x^{-2}-y^{-2}}.
\label{app:tail-estimator-bias}
\end{equation}
For $y=\lceil\lambda x\rceil$ with fixed $\lambda>1$, the
$\mathcal O(Jx^{-3})$ infinite-chain correction produces an
$\mathcal O(J/x)$ bias. Consequently,
\begin{equation}
\lim_{x\to\infty}\lim_{L\to\infty}
\widehat A_L(x,\lceil\lambda x\rceil)=A^{[4]}.
\label{app:tail-estimator-limit}
\end{equation}
The $L\to\infty$ limit uses commensurate rings at fixed $K,h,x,y$;
holding $y/L$ at a nonzero fraction is a different limit.
Keeping $y/x$ away from unity avoids the additional cancellation
associated with adjacent separations. Numerical errors in the row
sums are nevertheless amplified by the inverse denominator, which
scales as $x^2$ at fixed $\lambda$.

\begin{figure}[!tbp]
\centering
\includegraphics[]{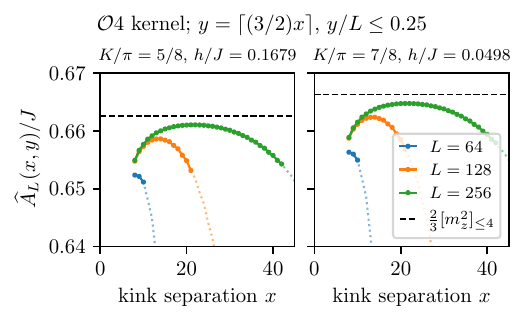}
\caption{Extraction of the dressed inverse-square tail from two
Hamiltonian row sums at $\alpha=4$.
The estimator $\widehat A_L(x,y)/J$ in
Eq.~\eqref{app:tail-estimator} is shown for
$K/\pi=5/8$, $h/J\simeq0.1679$ (left), and
$K/\pi=7/8$, $h/J\simeq0.0498$ (right).
Blue, orange, and green curves correspond to $L=64,128,256$,
respectively. Black dashed lines give the independent vacuum
prediction $(2/3)m_z^{2,[4]}$.
Solid curves mark $x\ge8$ and $y/L\le1/4$, with
$y=\lceil3x/2\rceil$; dotted continuations show larger separations.
Increasing $L$ brings the broad maximum closer to the prediction,
while finite-size bending persists at larger separations.}
\label{fig:kernel_tail_check}
\end{figure}

Figure~\ref{fig:kernel_tail_check} uses $y=\lceil3x/2\rceil$ and the
periodic-image coupling convention for $L=64,128,256$.
The horizontal prediction is
$A^{[4]}/J=(2/3)m_z^{2,[4]}$, where $m_z^{2,[4]}$ is the squared
bulk magnetization consistently expanded through fourth order and
obtained independently of the kernel extraction from the vacuum
calculation in Eqs.~\eqref{app:mz2B} and \eqref{app:mz2O4}.
It is a separate perturbative prediction, not a nonperturbative
magnetization measurement. At both displayed parameter pairs, increasing
$L$ from $128$ to $256$ brings the broad maximum closer to this
prediction and shifts it to larger separations; the $L=64$ trace is
already bending downward in the displayed range. The subsequent downward bending also moves
outward, identifying an intermediate range between short-distance
corrections and finite-ring effects.

These trends support finite-size consistency with the predicted tail.
The residual offset and curvature at $L=256$ show that the displayed
window has not reached its asymptotic limit. The selection $y/L\le1/4$
is geometric; controlled extraction requires stability with $L$ at fixed
separations, followed by increasing $x,y$ with $x,y\ll L$.
Resolving the small quartic contribution separately requires the
individual field-expansion coefficients. This comparison tests the
fourth-order effective theory. Unlike the correlation reconstruction
in Appendix~\ref{app:tail}, no plateau continuation fixes the estimator's
asymptotic amplitude.

\section{Scaling laws, remaining corrections, and conditional onset}
\label{app:scaling}
\label{app:diagnostics}

\subsection{Quantization, sizes, and counting}

Dividing Eq.~(4) by $D_K$ and keeping its two leading
terms gives the half-line equation
\begin{equation}
-f''(x)-\frac{g_K}{x^2}f(x)=-\kappa^2f(x),
\qquad \kappa^2=\frac{\delta}{D_K}.
\label{eq:scaleeq}
\end{equation}
At $\kappa=0$, substitution of $f=x^\lambda$ gives
$\lambda(\lambda-1)+g_K=0$. The two powers become
$1/2\pm\ii s_K$ above $g_K=1/4$. At the critical value they merge,
giving $\sqrt{x}$ and $\sqrt{x}\log x$; below it they are real.
For the fixed matching phase and bounded counting remainder used below,
we use the power-decaying potential remainder in Eq.~(3)
and assume derivative corrections whose contribution to the logarithmic
phase is integrable. The fourth-order kernel satisfies these conditions
with an $\cO(x^{-3})$ interaction
remainder and the hopping-moment bounds of Appendix~\ref{app:weakfieldO4}.
Under these conditions the infinite geometric hierarchy requires
$g_K>1/4$. Core-bound levels may also occur outside that regime.

For $g_K>1/4$, the decaying finite-binding solution is
$f_\kappa(x)=\sqrt{x}K_{\ii s_K}(\kappa x)$.
Its small-argument form is a sum of two logarithmic waves,
$K_{\ii s}(y)\sim[\Gamma(\ii s)(y/2)^{-\ii s}
+\Gamma(-\ii s)(y/2)^{\ii s}]/2$.
Matching their relative phase to a fixed core gives
\begin{equation}
s_K\log\frac1{\kappa_a x_{\rm core}}+\phi_K
=a\pi+o(1),
\label{app:quantization}
\end{equation}
at fixed $K,h$ as $a\to\infty$. Here $x_{\rm core}$ is a matching length
and $\phi_K$ is an energy-independent phase in the threshold limit.
It follows that $\kappa_{a+1}/\kappa_a\to\ee^{-\pi/s_K}$ and
$\delta_a=\delta_{\star,K}\ee^{-2\pi a/s_K}[1+o(1)]$, with a
core-dependent scale $\delta_{\star,K}$.

The norm of an increasingly shallow state is dominated by its outer
wave function. Consequently,
\begin{equation}
\begin{aligned}
\langle x^p\rangle_a&=\kappa_a^{-p}
\frac{I_p(s_K)}{I_0(s_K)}[1+o(1)],\\
I_p(s)&=\int_0^\infty y^{p+1}|K_{\ii s}(y)|^2\,dy.
\end{aligned}
\label{app:moments}
\end{equation}
For real $p>-2$, integration of the Bessel integral representation gives
\begin{equation}
I_p(s)=\frac{2^{p-1}\Gamma(1+p/2)^2
|\Gamma(1+p/2+\ii s)|^2}{\Gamma(p+2)}.
\label{app:Bessel_integral}
\end{equation}
In particular, $I_0(s)=\pi s/[2\sinh(\pi s)]$ and
$I_2(s)/I_0(s)=2(1+s^2)/3$. Thus
\begin{equation}
\xi_a=\frac{c_\xi(s_K)}{\kappa_a}[1+o(1)],\qquad
c_\xi^2(s)=\frac{2(1+s^2)}3,
\label{app:size_amplitude}
\end{equation}
and the binding--size product is
\begin{equation}
\delta_a\xi_a^2\longrightarrow
\frac{2D_K}{3}(1+s_K^2)=\frac{2A}{3}+\frac{D_K}{2}.
\label{app:binding_size}
\end{equation}
This evaluates the normalization-independent amplitude in addition to
the successive-level ratios.

Counting the allowed phases in Eq.~\eqref{app:quantization} gives
\begin{equation}
\cN_K(\delta)=\frac{s_K}{2\pi}
\log\frac{\delta_{\star,K}}{\delta}+\cO(1).
\label{app:energy_count}
\end{equation}
The leading coefficient agrees with the half-line discrete
inverse-square theorem~\cite{DamanikTeschl2007}: for a nearest-neighbor
discrete Laplacian and $V(n)=-g_K/n^2+w(n)$, a sufficient remainder
condition is $\sum_{n\geq1}n|w(n)|<\infty$.
That theorem fixes the logarithmic counting coefficient; it does not
by itself cover the full dressed hopping kernel or give the bounded
remainder used here, which follows from the limiting matching phase. For fixed $s_K$, the
infrared scale is proportional to $D_K/L^2$, with a coefficient that
depends on $s_K$ and the boundary condition. This yields
Eq.~(6). Counting is performed in one resolved channel:
combining independent parity sequences would change the multiplicity.
The resulting consistency relation can be stated without differentiating
a finite-size staircase,
\begin{equation}
\lim_{a\to\infty}\frac{2\pi}{\log(\delta_a/\delta_{a+1})}
=\lim_{a\to\infty}\frac{\pi}{\log(\xi_{a+1}/\xi_a)} =\lim_{L\to\infty}\frac{\pi\cN_K(L)}{\log L}
=\sqrt{\frac{2Jm_z^2}{3D_K}-\frac14}.
\label{eq:overconstrained}
\end{equation}
The counting law is an additional prediction beyond the energy and size
comparisons in the figures.

For fixed $s_K$, useful scaling diagnostics are
\begin{equation}
\begin{gathered}
\kappa_a x_{\rm core}\ll1,\qquad
\xi_a=\frac{c_\xi(s_K)}{\kappa_a}\ll L,\\
R_{4,a}=\frac{|C_K|\kappa_a^2}{D_K}\ll1,\qquad
R_{6,a}=\frac{|F_K|\kappa_a^4}{D_K}\ll1.
\end{gathered}
\label{app:controlsO4}
\end{equation}
The matching length must include the region where higher gradients
substantially modify the logarithmic wave function.
The explicit size condition matters when $s_K$ is large, since
$c_\xi(s_K)\propto s_K$: the weaker estimate $\kappa_a L\gg1$ alone
is then insufficient to place the entire meson well inside the ring.
The bare interaction has an $x^{-4}$ correction; the dressed-kernel
argument in Sec.~\ref{app:dtO4} only bounds the more general
interaction remainder by $\cO(x^{-3})$ through fourth order.
Quartic dispersion supplies the leading generic kinetic correction. Tuning $C_K=0$ suppresses the
quartic kinetic term; sextic dispersion and interaction-tail corrections
remain. The core phase and its energy dependence also govern how the
finite-level ratios approach their limits.

\subsection{Scattering and onset}

Above threshold write
$f_q=\sqrt{x}[A_+J_{\ii s_K}(qx)+A_-J_{-\ii s_K}(qx)]$.
A fixed real core condition fixes
$r_K(q)=A_-/A_+=\exp[2\ii s_K\log(q/q_\star)+\ii\theta_K]$,
where $\theta_K$ is real.
Using the large-argument Bessel forms, with incoming and outgoing waves
defined by $\ee^{\mp\ii(qx-\pi/4)}$, gives
\begin{equation}
S_K^{(0)}(q)=
\frac{\ee^{\pi s_K/2}+r_K(q)\ee^{-\pi s_K/2}}
{\ee^{-\pi s_K/2}+r_K(q)\ee^{\pi s_K/2}}.
\label{app:scattering_matrix}
\end{equation}
For real $q$, $|r_K|=1$ makes $|S_K^{(0)}|=1$, as required for an
isolated elastic channel. Since $r_K(\Lambda_K q)=r_K(q)$, the leading
matrix is periodic in $\log q$ with period $\pi/s_K$~\cite{HammerSwingle2006}. Corrections from the physical core and lattice
give $S_K(q)=S_K^{(0)}(q)+o(1)$, yielding
Eq.~(9). The periodic observable is the scattering
matrix; a particular choice of unwrapped phase need not be linear in
$\log q$.

For the momentum onset, define $\mathcal G(K)=A/D_K-1/4$ at fixed $h$.
The critical curvature is
\begin{equation}
D_{K_\star}=4A=\frac{8Jm_z^2}{3},
\label{eq:criticalDK}
\end{equation}
and $\mathcal G'(K_\star)=-D'_{K_\star}/(4D_{K_\star})$.
Parameterizing distance into the supercritical side by $\Delta K>0$
gives $\mathcal G(K)=B_\star\Delta K+\cO(\Delta K^2)$, with
$B_\star=|D'_{K_\star}|/(4D_{K_\star})$.
The universal ratios therefore have the essential singularities in
Eq.~(8). Their shallow-level limit is taken before
approaching $K_\star$.
For an individual branch, Eq.~\eqref{app:quantization} gives
\begin{equation}
\delta_a(K)=\frac{D_K}{x_{\rm core}^2}
\exp\!\left[-\frac{2(a\pi-\phi_K)}{s_K}\right][1+o(1)].
\label{app:absolute_onset}
\end{equation}
Thus the coefficient in an absolute branch law also depends on the
limiting core phase and the branch labeling. The extra logarithmic
oscillation between adjacent shallow states always contributes
$\ee^{-2\pi/s_K}$. The finite-$K$ onset is conditional on an actual
curvature crossing while the channel remains isolated. The
$C_K=0$ scan supplies a supercritical benchmark for the ratio laws.

\subsection{Departure from the marginal interaction}

Under the same assumption that dressing adds only subleading terms,
for $\alpha=4+\eta$ the tail is
$-A_\alpha x^{-2}\ee^{-\eta\log x}$, with
$A_\alpha=4Jm_z^2/[(\alpha-1)(\alpha-2)]$.
Away from $g_K=1/4$, approximate discrete scaling
requires $|\eta|\log(x/x_{\rm core})\ll1$.
One level per logarithmic size interval $\pi/s_K$ gives the estimate
$\cN_{\rm DSI}=\cO(s_K/|\eta|)$, further limited by ring and kinetic
effects. Its prefactor depends on the tolerated drift of ratios;
near $g_K=1/4$, the sensitivity of $s_K$ narrows the window.
For $\alpha>4$ the tail supports only finitely many bound levels;
for $2<\alpha<4$ the threshold accumulation is stronger than logarithmic
and is not asymptotically geometric.

\end{document}